%% file: Geometric_Invariants_of_Quantum_Metrology.tex
\documentclass[10pt,aps,pra,twocolumn,superscriptaddress,floatfix,nofootinbib]{revtex4-2}

\usepackage{amsmath, amsfonts, amssymb }
\usepackage{amsthm}

\usepackage{bm,bbm, braket, accents}
\usepackage{graphicx}   
\usepackage{soul}
\usepackage[usenames,dvipsnames]{xcolor}
\usepackage{lipsum} 
\usepackage{pgfplots}
\pgfplotsset{compat=newest}
\usepackage{comment}

\usepackage[normalem]{ulem}

\usepackage[breaklinks=true]{hyperref}
\hypersetup{
  colorlinks   = true, 
  urlcolor     = blue, 
  linkcolor    = blue, 
  citecolor    = red 
}

\usepackage{orcidlink}

\newcommand\Tr{\mathrm{Tr}}

\newcommand{\op}[2]{\ket{#1}\!\bra{#2}}

\newcommand{\mh}[0]{\mathcal{H}} 

\newcommand{\g}{\mathfrak{g}}
\newcommand{\G}{\mathfrak{G}} 
\newcommand{\M}{\mathcal{M}}
\newcommand{\su}[1]{\mathrm{su}(#1)}
\newcommand{\uu}[1]{\mathrm{u}(#1)}
\newcommand{\SU}[1]{\mathrm{SU}(#1)}
\newcommand{\F}{\mathcal{F}}
\DeclareMathAlphabet{\numbb}{U}{BOONDOX-ds}{m}{n}
\newcommand{\bh}{\mathcal{D}(\mathcal{H})} 
\newcommand{\Ubb}{\mathbb{U}}

\usepackage[bb=boondox]{mathalfa}

\usepackage{multirow}
\usepackage{floatrow}

\usepackage[normalem]{ulem}

\usepackage[capitalise]{cleveref} 
\crefformat{equation}{Eq.~(#2#1#3)} 
\crefformat{section}{Sec.~#2#1#3} 
\Crefformat{equation}{Equation~(#2#1#3)}
\crefformat{figure}{Fig.~#2#1#3}
\crefrangeformat{equation}{Eqs.~#3(#1)#4--#5(#2)#6}
\Crefformat{section}{Section~#2#1#3}

\usepackage{graphicx}

\begin{document}

\title{Geometric Invariants of Quantum Metrology}

\author{Christopher Wilson\orcidlink{0000-0002-7609-2189}}
\thanks{Corresponding author: Chris.Wilson-2@colorado.edu}
\affiliation{JILA, Department of Physics, University of Colorado, Boulder, Colorado 80309, USA}
\affiliation{\normalfont{These authors contributed equally to this work.}}

\author{John Drew Wilson\orcidlink{0000-0001-6334-2460}}
\affiliation{JILA, Department of Physics, University of Colorado, Boulder, Colorado 80309, USA}
\affiliation{\normalfont{These authors contributed equally to this work.}}

\author{Luke Coffman\orcidlink{0000-0003-1539-6418}}
\affiliation{JILA, Department of Physics, University of Colorado, Boulder, Colorado 80309, USA}

\author{Shah Saad Alam\orcidlink{0000-0001-6541-662X}}
\affiliation{JILA, Department of Physics, University of Colorado, Boulder, Colorado 80309, USA}

\author{Murray J. Holland\orcidlink{0000-0002-3778-1352}}
\affiliation{JILA, Department of Physics, University of Colorado, Boulder, Colorado 80309, USA}

\date{\today}


\begin{abstract}
We establish a previously unexplored conservation law for the Quantum Fisher Information Matrix (QFIM) expressed as follows; when the QFIM is constructed from a set of observables closed under commutation, i.e., a Lie algebra, the spectrum of the QFIM is invariant under unitary dynamics generated by these same operators.
Each Lie algebra therefore endows any quantum state with a fixed “budget” of metrological sensitivity---an intrinsic resource that we show, like optical squeezing in interferometry, cannot be amplified by symmetry‑preserving operations.
The Uhlmann curvature tensor (UCT) naturally inherits the same symmetry group, and so quantum incompatibility is similarly fixed.
As a result, a metrological analog to Liouville's theorem appears; statistical distances, volumes, and curvatures are invariant under the evolution generated by the Lie algebra.
We discuss this as it relates to the quantum analogs of classical optimality criteria.
This enables one to efficiently classify useful classes of quantum states at the level of Lie algebras through geometric invariants.
\end{abstract}


\maketitle

Conservation laws are of fundamental importance to physics, and it has been long known that conserved quantities corresponds to an underlying symmetry~\cite{Noether1971}.
Here, we derive a conservation law for the Quantum Fisher Information (QFI) and Uhlmann Curvature Tensor (UCT) from the underlying symmetries of quantum state space.

The QFI is a fundamental object in quantum mechanics, where it serves as an entanglement witness ~\cite{Liu2019,Hyllus}, a generalized resource witness~\cite{Tan,Kudo2023}, plays a fundamental role in quantum metrology~\cite{Pezze2018_NonclasAtoms,Wilson2023} and quantum state geometry~\cite{Braunstein1994,Wootters,Wootters,Campos}, and is a signature of quantum phase transitions~\cite{Wang2014,Wu2016,Song2017,Chiafalo2019, Cieslinski2024, Fang2025}.
When extended to multiple parameters, the QFI Matrix (QFIM) serves as a Riemannian metric on the space of quantum states~\cite{Petz1999,Sidhu,Paivi}.
Similarly, the UCT describes the curvature of quantum state space or, equivalently, the metrological incompatibility between parameters due to measurement induced back action~\cite{Carollo2019}. 
This is the mixed state generalization of the Berry curvature~\cite{Uhlmann1993} and it has proven important to quantum phase transitions~\cite{Carollo,Viyuela,Viyuela2,Hou,Galindo}, quantum steering~\cite{Yao} and, of course, quantum metrology~\cite{Carollo2019,Wilson2023,Guo}.
These two geometric quantities are witnesses to key quantum resources for quantum technologies which surpass their classical counterparts~\cite{NielsenChaung,Shor,Kendon,Bennett2,Yin,Holland1993,Zoller2024,Buscemi}.

The efficient characterization of these quantities is a task of broad applicability and interest~\cite{Beckey2021,Beckey2022,Kang2025,Barnes,Yu,Yu2,alam2024}.
In particular, understanding when the QFI can be increased by introducing a quantum resource (like entanglement) is necessary for quantum advantage in sensing~\cite{Kitagawa1993,Toth2,Greve} while engineering robust quantum processes often comes down to preventing the QFI from being adversely modified~\cite{Meyer2021,Reilly2024} by decoherence.
Similarly, useful quantum multi-parameter metrology can often be reduced to minimizing the UCT between parameters of interest~\cite{Liu2019,Zoller2023,Wilson2023}. 
Characterizing the set of dynamics which either protect or modify these tensors is indispensable to both tasks.

\begin{figure}[t]
    \centering
    \includegraphics[width= 1 \columnwidth]{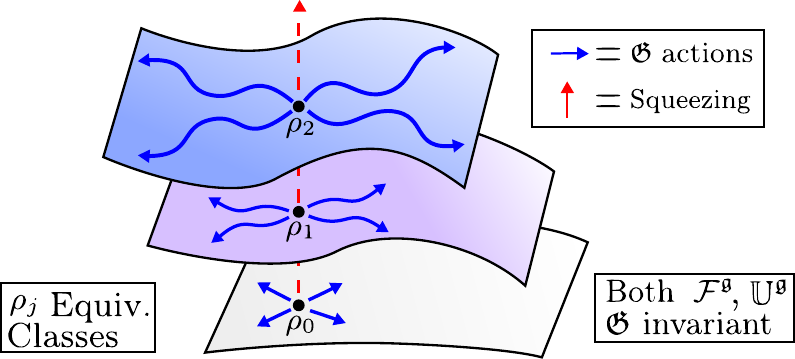}
    \caption{
    The QFIM, $\mathcal{F}^\g$, and UCT, $\mathbb{U}^\g$, built from a Lie algebra, $\g$, are invariant to $\G$-actions (unitaries generated by $\g$, blue solid arrows).
    This gives rise to equivalence class manifolds where actions not in $\G$ (red dashed arrows) move between these classes.
    Here, squeezing modifies the QFIM spectrum and UCT so states evolve faster and with modified curvature (growing and bending the blue arrows). 
    }
    \label{fig:Cartoon}
\end{figure}

We establish a simple conservation law to aid in this characterization;
when one constructs the QFIM from a set of observables closed under commutation, the eigenvalues of the QFIM are invariant under the unitary evolution generated by these same observables.
Furthermore, we show that the UCT inherits this same symmetry group.
Every such set of observables, or Lie algebra, therefore endows any quantum state with a fixed “budget” of metrological sensitivity and incompatibiltiy---intrinsic resources that we show, like optical squeezing~\cite{Braunstein2005}, cannot be modified by symmetry‑preserving operations.
We connect this to several optimality criteria~\cite{kiefer1959} based on functionals of the QFIM.
This equates the general ``usefullness" of a quantum state under a resource theory~\cite{Tan,Kudo2023} to symmetries at the level of Lie groups and algebras (summarized by~\cref{fig:Cartoon}).
These symmetry-preserving operations partition the quantum state space into equivalence classes, where each class is characterized by geometric objects such as the spectrum of the QFIM and the meteorological incompatibility.

These results lead to a metrological analog to Liouville's theorem~\cite{Liouville1838}.
When these geometric objects are constructed from an underlying Lie algebra, the equivalence classes that emerge are manifolds. On these manifolds, the symmetry group does not stretch distances, given by the spectrum of QFIM, nor compress volume, given by determinant of the QFIM, nor curve state space, given by elements of the UCT, nor introduce quantum incompatibilities, given by the metrological incompatibility parameters.
These quantities and any functionals thereof are geometric invariants for any quantum metrological task.

\section{Formalism}
We wish to characterize a probe state, $\rho$, on a finite dimensional Hilbert space, $\mh$, through the lens of the QFIM, which is defined element-wise by
\begin{equation}\label{eq:QFIMdef}
    \F[\rho]_{\mu \nu} \equiv \frac{1}{2} \Tr_{\mh}[ \rho \{ \hat{L}_\mu, \hat{L}_\nu \} ],
\end{equation}
where $\{\hat{A},\hat{B}\} = \hat{A} \hat{B} + \hat{B} \hat{A}$, and $\hat{L}_\mu$ is the symmetric logarithmic derivative (SLD)~\cite{Braunstein1994,Liu2019} with respect to the parameter $\theta^\mu$. 
While $\rho$ may be a mixed or pure state, we only consider unitary parameter encoding.
The SLD $\hat{L}_\mu$ is thus implicitly defined through the equation
\begin{equation} \label{eq:SLDDef}
\frac{\partial \rho}{\partial \theta^\mu}  = -i [ \hat{G}_\mu, \rho] = \frac{1}{2} \{ \hat{L}_\mu, \rho \},
\end{equation} 
where $[\hat{A},\hat{B}] = \hat{A} \hat{B} - \hat{B} \hat{A}$, and $\hat{G}_\mu$ is the Hermitian observable that generates evolution in the $\theta^\mu$ direction.
Throughout, $\hbar = 1$ and repeated upper-lower index implies summation.

To establish a conservation law on the QFIM, we will choose the set of Hermitian operators $\{ \hat{G}_{\mu}\}$ such that they constitute a Lie algebra $\g$.
This means we unitarily encode some parameter $\omega$ via the Hamiltonian 
\begin{equation}
\label{eq:eigvec}
\hat{H}[\vec{V}] \equiv  \omega \sum_{\mu} V^\mu \hat{G}_\mu,
\end{equation}
with each $\hat{G} \in \g$.
In order to ensure that the eigenvectors of the QFIM correspond to statistically independent directions of evolution in Hilbert space~\cite{Wilson2023}, we may also constrain the set $\{\hat{G}_{\mu}\}$ to be mutually orthogonal under the Hilbert Schmidt inner product \cite{suppMatCons}. 
When the QFIM is built on such a basis $\{\hat{G}_{\mu}\}$, the components will be denoted as $\mathcal{F}^{\g}[\rho]_{\mu \nu}$ with matrix representation $\mathbf{F}^{\g}[\rho]$. The Hamiltonian in~\cref{eq:eigvec} is chosen such that 
\begin{equation}\label{eq:EigVec}
\mathbf{F}^{\g}[\rho] \vec{V} = \lambda \vec{V}.
\end{equation}
The estimation of the parameter $\omega$ is therefore lower bounded by a precision of
\begin{equation}
\Delta \omega^2 \geq \frac{1}{M t^2 \big|\vec{V}\big|^2 \lambda }
\label{eq:QCRB}
\end{equation}
for $M$ many independent trials each with an interrogation time of $t$~\cite{Paris2009,Liu2019,Pezze2018_NonclasAtoms} and where $|\vec{V}|^2 = \vec{V} \cdot \vec{V}=V^{\mu} \delta_{\mu \nu} V^{\nu}$.
Eigenvectors and Eigenvalues of the QFIM effectively serve as a principal component analysis for $\rho$ under $\g$-generated evolution.
With these relatively simple definitions in place, we can now state the main result.

\section{Theorem}
\noindent \textit{The eigenvalues of a quantum Fisher information matrix, $\F^{\g}[\rho]_{\mu \nu}$, constructed from an orthonormal basis of observables constituting a Lie algebra $\g$, are invariant under unitary evolution generated by $\g$---the Lie group, $\G$.}

\vspace{\baselineskip}

\noindent \textbf{Proof:} We want to compare the QFIM for the probe state $\rho$ to the QFIM for the unitarily evolved state 
\begin{equation}
\rho\mapsto \tilde{\rho}=\hat{U} \rho \hat{U}^\dagger
\label{eq:transform}
\end{equation}
for unitary operator $\hat{U}$.
To do so, we can use the following intuition: if each SLD is identified by a derivative, and each derivative is given by a commutator of the state with a Hermitian observable, then the SLD transforms the same as these observables under $\hat{U}$.
As a result, the two QFIMs will be related by a simple transformation law which, we will show, doesn't change the spectrum of the QFIM as long as $\hat{U}$ is generated by the same set of Hermitian observables used to identify the SLDs.
The cyclicity of the trace in the definition of the QFIM gives;
\begin{equation} 
\begin{aligned} \label{eq:QFIMCycle}
    \F^{\g}[\tilde{\rho}]_{\mu \nu} &= \frac{1}{2} \Tr_{\mh} [ \tilde{\rho} \{ \tilde{L}_\mu, \tilde{L}_\nu \}] \\
    &= \frac{1}{2} \Tr_{\mh} [ \rho \{ \hat{U}^\dagger \tilde{L}_\mu \hat{U}, \hat{U}^\dagger \tilde{L}_\nu \hat{U} \} ],
\end{aligned}
\end{equation}
To establish a transformation law between the $\{L_{\mu}\}$ and $\{\tilde{L}_{\mu}\}$, we need three things. 
First, the operators $\{ \hat{G}_{\mu}\}$ associated to the SLDs must form a Lie algebra $\g$.
Second, the unitary $\hat{U}$ must belong to the Lie group $\G$ generated by $\g$.
Lastly, we require that our basis of $\g$ is normalized with respect to the Hilbert-Schmidt norm such that $\Tr_{\mh}(\hat{G}_\mu \hat{G}_\nu) = C \delta_{\mu\nu}$ for the norm $C$ and $\delta_{\mu\nu}$ the Kronecker delta function.

These conditions cause the QFIM to inherit transformation behavior from the Lie group $\G$:
\begin{equation} \label{eq:conjugate}
\hat{U}^\dagger \tilde{L}_\mu U = \Lambda^\alpha_\mu \hat{L}_\alpha, \quad \mathrm{and} \quad \F^{\g}[\tilde{\rho}]_{\mu\nu
} = \Lambda^\alpha_\mu \Lambda^\beta_\nu \F[\rho]_{\alpha\beta},
\end{equation}
where $\Lambda^\alpha_\mu$ are the matrix elements in the $\{G_{\mu}\}$ basis of the linear map $\mathbf{\Lambda}$ which implements the adjoint action $\mathrm{Ad_{U^{\dagger}}}: \g \to \g$. 
Our orthonormality condition ensures $\mathbf{\Lambda} \mathbf{\Lambda}^T = \mathbf{\Lambda}^T \mathbf{\Lambda} = \mathbf{1}$, forcing $\mathbf{\Lambda}\in \mathrm{SO}(\dim \g)$.

On the surface, this is similar to the re-parameterization law for the QFIM~\cite{Liu2019}. 
The difference is that here the state $\rho$ has evolved to $\tilde{\rho}$ and the parameterization, i.e. the basis $\{\hat{G}_\mu\}$, has remained fixed. 
Our assumptions guarantee~\cref{eq:transform} preserves the structure of $\g$ and therefore is an isometry of $\mathcal{F}^{\g}$.
In matrix form, $\mathbf{F}[\tilde{\rho}] = \mathbf{\Lambda}^{T} \mathbf{F}[\rho] \mathbf{\Lambda}$, and thus
\begin{equation}
\begin{aligned}
\mathbf{F}^{\g}[\rho] \vec{V} = \lambda \vec{V}, \quad \mathrm{and} \quad
\mathbf{F}^{\g}[\tilde{\rho}] \left( \mathbf{\Lambda}^{T} \vec{V}\right)=& \lambda \left(\mathbf{\Lambda}^{T} \vec{V} \right).
\end{aligned}
\end{equation}
This means $\mathbf{F}^{\g}[\tilde{\rho}]$ and $\mathbf{F}^{\g}[\rho]$ share the same eigenvalues---the same spectrum.
Since $\hat{U}$ was an arbitrary unitary operator in $\G$, the eigenvalues of the QFIM are invariant to all of $\G$. $\blacksquare$

In Ref.~\cite{Wilson2025} we show the same invariance holds in a $\g$-basis independent manner to formally make the result geometrically well phrased, and we extend this invariance to arbitrary monotone metrics~\cite{Petz1996} and to the normalizer of $\G$ in the full Unitary group over $\mh$.
Additionally, we show that on infinite Hilbert spaces this invariance is significantly modified. 
The mathematical machinery necessary for these results falls outside the scope of discussion here.

\subsection{Lemma: The Uhlmann Curvature and Metrological Incompatability}
As a lemma to our main theorem, we show that the UCT built from $\g$ inherits the same symmetry, where~\cite{Liu2019,Carollo,Candeloro}
\begin{equation}\label{eq:UCTdef}
    \Ubb^\g[\rho]_{\mu \nu} \equiv -\frac{i}{2} \Tr_{\mh}\left[ \rho [ \hat{L}_\mu, \hat{L}_\nu ] \right].
\end{equation}
Physically, the UCT computes the quantum incompatibility of simultaneous estimations due to the underlying non-commutativity,
\begin{equation}
    \vec{V}^T \mathbf{U}^{\g}[\rho] \vec{W} =  V^\mu\Ubb^{\g}[\rho]_{\mu \nu} W^\nu.
\end{equation}
When $\vec{V}^T \mathbf{U}^{\g}[\rho] \vec{W} = 0$, one can simultaneously and optimally estimate the two frequencies of the Hamiltonian: $\hat{H} = \omega V^\mu \hat{G}_\mu + \eta W^\nu \hat{G}_\nu$. We now state the lemma:

\vspace{\baselineskip}

\noindent \textbf{Lemma:} \textit{The eigenvectors of the quantum Fisher information matrix constructed from the orthonormal basis of $\g$ have invariant mutual Uhlmann curvature under $\G$.}

\vspace{\baselineskip}

\noindent \textbf{Proof:} Consider $\mathbf{F}^\g[\rho]$ and $\mathbf{F}^\g[\tilde{\rho}]$, where again, $\tilde{\rho}=\hat{U} \rho \hat{U}^{\dagger}$ for some $\hat{U} \in \G$. Let $\vec{V}$ and $\vec{W}$ be eigenvectors of $\mathbf{F}^\g[\rho]$, which, per the main theorem, specify eigenvectors of $\mathbf{F}^\g[\tilde{\rho}]$; $\mathbf{\Lambda}^T \vec{V}$ and $\mathbf{\Lambda}^T \vec{W}$ respectively.
By the cyclicity of the trace $\mathbf{U}^{\g}[\tilde{\rho}] = \mathbf{\Lambda}^T \mathbf{U}^{\g}[\rho] \mathbf{\Lambda}$ and therefore:
\begin{equation}
\begin{aligned}
    \vec{V}^T \mathbf{U}^{\g}[\rho] \vec{W} = \left( \mathbf{\Lambda}^T \vec{V}\right)^T \mathbf{U}^{\g}[\tilde{\rho}] \left( \mathbf{\Lambda}^T \vec{W} \right) \quad \blacksquare
\end{aligned}
\end{equation}

We can use this Lemma to establish that the metrological incompatibility is invariant as well, but first we provide some geometric intuition.
Every metric defines an inner product, and the QFIM is no different; $\langle \vec{V}, \vec{W} \rangle_{\F^\g} \equiv \vec{V}^T \mathbf{F}^\g[\rho]\vec{W}$.
This diagnoses quantum statistical independence~\cite{Suzuki,Alam} between parameters unitarily encoded by $\vec{V}$ and $\vec{W}$ via~\cref{eq:eigvec} the same way the Euclidean inner product diagnoses orthogonality between vectors in real space.

The UCT, while not a metric, similarly acts on two vectors and outputs a scalar that diagnoses quantum incompatibility.
Tensors like these, with two lower indices, don't \textit{formally} have well defined eigenvalues and eigenvectors~\cite{footnoteMetric}.
To remedy this, a separate metric must be used to ``raise an index''. Previously, this was done implicitly when we used the Hilbert-Schmidt inner-product to construct our orthornormal basis $\{G_{\mu}\}$ of $\g$, allowing us to raise an index in a geometrically well defined way with the Kronecker delta $\delta^{\mu \nu}$.

\begin{figure}[t]
    \centering
    \includegraphics[width= 1 \columnwidth]{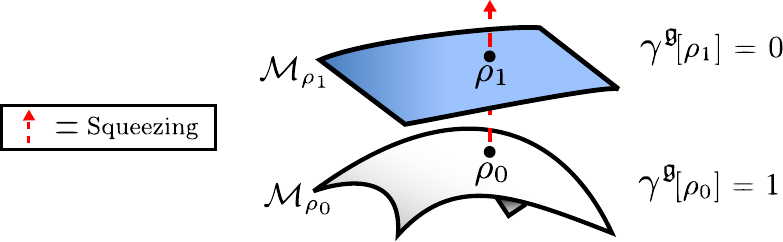}
    \caption{ A representation of the changing incompatibility under an action not in $\G$, labeled abstractly as squeezing.
    }
    \label{fig:FigCurv}
\end{figure}

For the UCT, the (Moore-Penrose) inverse of $\F^{\g}[\rho]_{\mu \nu}$ does the same, thereby guaranteeing well-defined eigenvalues and eigenvectors.
The raised index UCT is then 
\begin{equation}
\mathbb{U}^{\g}[\rho]^{\mu}_{\ \nu} \equiv \mathcal{F}^{\g}[\rho]^{\mu \alpha} \mathbb{U}^{\g}[\rho]_{\alpha \nu},
\end{equation}
where we use the convention that $\mathcal{F}^{\g}[\rho]^{\mu \alpha}$ with upper indices is the (Moore-Penrose) inverse of $\F^{\g}[\rho]_{\mu \alpha}$. The eigenvalues of the raised index UCT account for the degradation of attainable precision due to non-commutativity, and the largest eigenvalue defines the metrological incompatibility parameter~\cite{Carollo2019, footnoteIncompatibility}
\begin{equation}
\begin{aligned}
\gamma[\rho] =& || i (\mathbf{F}^{\g}[\rho])^{-1} \mathbf{U}^{\g} [\rho] ||_\infty,
\label{eq:Incompatibility}
\end{aligned}
\end{equation}
where $||\mathbf{A}||_\infty$ is the largest eigenvalue of $\mathbf{A}$. 
Because both $\mathbf{F}^\g$ and $\mathbf{U}^{\g}[\rho]$ transform covariantly with $\G$, the spectrum of $\mathbb{U}^{\g}[\rho]^{\mu}_{\ \nu}$ is similarly invariant on $\M_\rho$. Thus, the metrological incompatibility, just like squeezing, is an irreducible resource.
We show an intuitive picture of the incompatibility in~\cref{fig:FigCurv}.

Ref.~\cite{Carollo2019} shows that the eigenvalues of $\mathbb{U}^{\g}[\rho]^{\mu}_{\ \nu}$ come in positive-negative pairs, and are bounded by $\pm1$. 
A largest eigenvalue of $0$ corresponds to vanishing Uhlmann curvature $\mathbb{U}^{\g}_{\mu \nu}$, and thus all parameters encoded by $\g$ may be optimally measured simultaneously. 
Notably, this means that the ratio between the Holevo bound~\cite{Holevo} and the Quantum Cramer Rao bound is one~\cite{Helstrom1969}.
Conversely, a largest eigenvalue of $1$ indicates maximum incompatibility, meaning~\cref{eq:QCRB} cannot be simultaneously saturated for multiple parameter.

\begin{table*}[t]
\begin{tabular}{ |c|l|c|l|c| } 
\hline
Opt.
& Classical Meaning 
& Quantum Form 
& Geometric Meaning 
& Geometric Form 
\\
\hline
\multirow{2}{3em}{A}
& \multirow{2}{8.2em}{Average variance
} 
& \multirow{2}{7.2em}{$\Tr_{\g}(\mathbf{F}^{\g}[\rho])$
} 
& \multirow{2}{6.2em}{Average $r^2$ of the $\epsilon$-ball} 
& \multirow{2}{8.2em}{$\overline{r^2_\g} = \frac{\Tr_{\g}(\mathbf{F}^\g[\rho])}{\dim(\g)} \frac{\epsilon^2}{8}$
} 
\\ & & & & \\
\multirow{2}{3em}{C} 
& \multirow{2}{8.2em}{Single parameter variance
} 
& \multirow{2}{7.2em}{$\vec{V}^T \mathbf{F}^{\g}[\rho] \vec{V}$
}  
& \multirow{2}{6.2em}{Arc length along $\vec{V}$
} 
& \multirow{2}{12.2em}{$\mathrm{d}s_{\vec{V}} = \sqrt{ \vec{V}^T \mathbf{F}^{\g}[\rho] \vec{V} } \mathrm{d}t$
}
\\& & & & \\ 
\multirow{2}{3em}{D} 
& \multirow{2}{8.2em}{Best generalized variance
} 
& \multirow{2}{7.2em}{$\mathrm{pdet}(\mathbf{F}^{\g}[\rho])$
} 
& \multirow{2}{6.2em}{Volume form
} 
& \multirow{2}{14.7em}{$\mathrm{dVol} = \sqrt{ \mathrm{pdet}(\mathbf{F}^{\g}[\rho]) } \ \prod_{\lambda_j\neq0} dx^j$
} 
\\ & & & & \\
\multirow{2}{3em}{E}
& \multirow{2}{8.2em}{Smallest eigenvalue
} 
& \multirow{2}{7.2em}{$\min_i[\lambda_i]$
} 
& \multirow{2}{6.2em}{Smallest direction} 
& \multirow{2}{14.7em}{$\mathrm{d}s_\mathrm{min} = \min_{|\vec{V}|^2=1} ( \mathrm{d}s_{\vec{V}} )$
} 
\\ & & & & \\
\multirow{2}{3em}{S} 
& \multirow{2}{8.2em}{Column orthogonality} 
& \multirow{2}{7.2em}{ $\frac{\mathrm{pdet} \mathbf{F}^\g[\rho]}{\prod_{\mathrm{diag}_j\neq0} \mathrm{diag}_j} $ } 
& \multirow{2}{7.2em}{Parameter-alignment
} 
& \multirow{2}{14.7em}{$\g$-basis dependent
}
\\ & & & & \\
\multirow{2}{3em}{T} 
& \multirow{2}{8.2em}{Distinguishability of two states} 
& \multirow{2}{7.2em}{same as Geometric} 
& \multirow{2}{6.2em}{Distance
} 
& \multirow{2}{14.7em}{$\Delta^2(\rho,\tilde{\rho}) = 1 - \Tr_{\mh}\left((\sqrt{\rho} \tilde{\rho}\sqrt{\rho} )^{1/2}\right) $
}
\\ & & & & \\
\hline
\end{tabular}
\caption{ Selected optimality criteria. A, D, and E-optimality are always $\G$-invariant. 
C-optimality can be further optimized over the state $\rho\in\M_\rho$, where it is $\G$-invariantly upper bounded by the largest eigenvalue of $\mathbf{F}^\g[\rho]$.
S-optimality is basis dependent and therefore never invariant. In the quantum form, $\mathrm{diag}_j$ is the $j^\mathrm{th}$ diagonal element of the QFIM.
T-optimality is geometrically invariant between states which co-evolve, i.e. if $\rho$ and $\tilde{\rho}$ evolve with the same unitary. 
In C-optimality, maximizing $\rho$ for a given $\vec{V}$ is often the goal of single parameter quantum sensing, and optimizing $\Vec{V}$ given $\rho$ was done in ref.~\cite{Wilson2023}.
}\label{tab:Criteria}
\end{table*}
%
\section{Equivalence Class Manifolds}
Any set of closed observables naturally carries with it a foliation (i.e. a partition into manifolds) of the set of all density operators, denoted $\bh$.
Thus, any Lie algebra can be used to partition $\bh$ into the equivalence classes given by the $\G$-adjoint orbits of $\rho$:
\begin{equation}\label{eq:EquivClass}
\mathcal{M}_\rho = \left\{ \tilde{\rho} \ \Big| \tilde{\rho} = \hat{U} \rho \hat{U}^\dagger \ \mathrm{for} \ \hat{U}\in \G \right\}.
\end{equation}
Our theorem states that, on these equivalence classes, the spectrum of $\mathbf{F}^\g[\rho]$ is fixed.
This leads to an immediate geometrically flavored insight: the manifold dimension of $\M_{\rho}$ is the number of non-zero eigenvalues of $\mathbf{F}^\g[\rho]$.
Moreover, this dimension is the maximum number of statistically independent parameters encoded by $\G$ one could hope to estimate from measurements of $\rho$. Lastly, in the case that $\gamma^{\g}[\rho]=0$, then one may simultaneously estimate all such parameters optimally. Our proof of invariance extends this optimality to \emph{any} state in $\mathcal{M}_{\rho}$.

An immediate consequence is that if one wishes to use a variational algorithm or similar method in order to optimize the QFIM with respect to a Lie algebra, the search space of states may be reduced by modding out by these equivalence classes. This can be phrased through the lens of quantum resource theories; every Lie algebra foliates $\bh$ into equivalence classes of states with the same ``resourcefulness". Actions in $\G$ cannot change this metrological budget, and therefore need not be searched over.

\subsection{Optimality Criteria}

More broadly, since each eigenvalue of $\mathbf{F}^{\g}[\rho]$ is conserved individually, \textit{any} functional of the QFIM's spectrum is invariant.
Optimizing such functionals of the Classical Fisher Information Matrix (CFIM) is often the goal of classical statistics for optimal experimental design~\cite{kiefer1959}. In the quantum case, any state $\rho_{\mathrm{Opt}}$ that is optimal under a given criteria automatically fixes that criteria on the entire equivalence class $\M_{\rho_{\mathrm{Opt}}}$. Accordingly, each criteria has a geometric meaning. 
We give a (by no means exhaustive) table of criteria derived from the QFIM and the physical interpretation in Table~\ref{tab:Criteria}. 
We note that any criteria with explicit vector dependence will not be geometrically invariant, but will be upper bounded on each $\M_{\rho}$ by a geometric invariant.

Two of the most common criteria are that of A-optimality (the trace over $\g$) and D-optimality (the determinant of non-zero eigenvalues):
\begin{equation} \label{eq:TrAndDet}
\Tr_{\g}(\mathbf{F}^{\g}[\rho]) = \sum_{i} \lambda_i, \ \  
\mathrm{pdet}(\mathbf{F}^{\g}[\rho]) = \prod_{\lambda_i \neq 0} \lambda_i
\end{equation} 
where $\lambda_{i}$ is the $i^{\mathrm{th}}$ eigenvalue, and the pseudo-determinant removes the null contributions of $\g$.

\begin{figure}
    \centering
    \includegraphics[width= 0.85 \columnwidth]{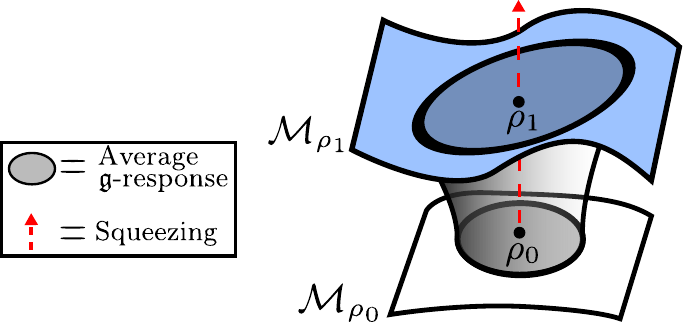}
    \caption{ A representation of two equivalence classes with different average $\g$-responses representing the A-optimality criteria. These change under actions not in $\G$, again labeled abstractly as squeezing.
    }
    \label{fig:FigAOpt}
\end{figure}

The geometric meaning of the traced QFIM is found by considering a ``small" unitary applied to the probe state; $\hat{U}(\epsilon,\vec{n}) = \mathrm{e}^{-i \epsilon \hat{G}_{\vec{n}} }$, where $\epsilon\ll1$, $\vec{n}$ is a $\dim(\g)$ unit vector such that $\hat{G}_{\vec{n}} = n^\mu \hat{G}_\mu \in \g$.
These unitaries constitute the ``$\epsilon$-ball'' around the identity of $\G$.
We can compare this to the set of states $\tilde{\rho}(\vec{n}) = \hat{U}(\epsilon,\vec{n}) \rho \hat{U}(-\epsilon,\vec{n}) $ in $\M_\rho$ which, unlike the $\epsilon$-ball in $\G$, is not spherical because $\rho$ does not isotropically respond to $\g$.
We can still compute the average squared distance from $\rho$ to $\tilde{\rho}(\vec{n})$, yielding the average response of $\rho$ to $\g$.
We do this via the squared Uhlmann distance~\cite{Hayashi2017,Liu2014,Beckey2021,Sidhu},
\begin{equation}
\begin{aligned}
\Delta^2 =& 1 - \Tr_{\mh}\left(\sqrt{\sqrt{\rho} \tilde{\rho}(\vec{n}) \sqrt{\rho} }\right) \approx \ \vec{n}^T \mathbf{F}^\g[\rho] \vec{n} \ \frac{\epsilon^2}{8}.
\end{aligned}
\end{equation}
Integrating this around the surface of the $\dim(\g)$-sphere~\cite{footnoteHaar,suppMatCons} givess the average squared distance:
\begin{equation}
\begin{aligned} \label{eq:AvgQFI}
\overline{r_\g^2} \equiv \frac{1}{A} \int_{S^{\dim(\g)}} \Delta^2 dA = \frac{\Tr_{\g}(\mathbf{F}^\g[\rho])}{\dim(\g)} \frac{\epsilon^2}{8},
\end{aligned}
\end{equation}
where $A$ is the surface area of the $\dim(\g)$-sphere and $dA$ is the area element, both given in the SM~\cite{suppMatCons}.
This provides a $\G$-invariant coarse graining of a state response to a given Lie algebra, shown conceptually in~\cref{fig:FigAOpt}.

We can upper bound this average response for an arbitrary Lie algebra, and evaluate it exactly for the case of $N$ many qudits, or $d$-level atoms, where the Lie algebra of interest is $\g = \su{d}$.
The QFI is maximal on pure states, where we can use the simplified expression~\cite{Liu2019}
\begin{equation}\label{eq:QFIMpure}
\mathcal{F}[\rho]_{\mu\nu} = 2 \langle \{\hat{G}_\mu,\hat{G}_\nu \}\rangle_\rho - 4 \langle \hat{G}_\mu \rangle_\rho \langle \hat{G}_\nu \rangle_\rho
\end{equation}
to find the trace
\begin{equation}
\Tr_{\g}( \mathbf{F}^\g[\rho] ) = 4 \sum_\mu \langle \hat{G}_\mu^2 \rangle_\rho - \langle \hat{G}_\mu \rangle_\rho^2.
\end{equation}
We can identify $\hat{\mathcal{G}}^2 \equiv \sum_{\mu} \hat{G}_\mu^2$ as the quadratic Casimir operator---the generalization of the dipole length from $\su{2}$.
If $\g$ is an irreducible representation, then $\hat{\mathcal{G}}^2$ has one distinct eigenvalue on $\mh$.
Whereas if $\g$ is a reducible representation, the optimal state will be in the subspace of $\mh$ with the largest Casimir eigenvalue. 
This is called the highest weight sub-representation. In both cases, we denote this largest eigenvalue by $\zeta$:
\begin{equation}
\Tr_{\g}( \mathbf{F}^\g[\rho] ) = 4 \zeta - 4 \sum_{\mu} \langle \hat{G}_\mu \rangle_\rho^2.
\end{equation}
This leads to a practical conclusion; to experimentally measure the A-optimality criteria of a pure state, one needs only $\mathrm{dim}(\g)$ many first order expectation values $\langle G_{\mu} \rangle_{\rho}$. Any quantum A-optimal state for $\g$ parameter sensing now satisfies
\begin{equation}\label{eq:AoptState}
\rho_\mathrm{A\text{-}Opt} =\underset{\rho}{\mathrm{argmin}} \sum_\mu \langle \hat{G}_\mu \rangle_\rho^2
\end{equation}
Intuitively, a state with a high average response must democratically minimize the expectation value of orthonormal operators that span $\g$.

The quantum D-optimality criteria similarly has a geometric interpretation, the volume form on $\mathcal{M}_\rho$:
\begin{equation}
\mathrm{dVol} = \sqrt{ \mathrm{pdet}(\mathbf{F}^{\g}[\rho]) } \ \prod_{\lambda_j\neq0} dx^j,
\end{equation}

where each coordinate $x^{j}$ on $\M_{\rho}$ is unitarily generated by $V_{(j)}^{\mu} \hat{G}_{\mu}$, with $V_{(j)}^{\mu}$ corresponding to the components of the $j^\mathrm{th}$ non-zero eigenvector of the QFIM. 
This volume form is also called the quantum Jeffrey's prior, the quantum analog of the invariant Bayesian prior for all unique parameterizations of $\mathcal{M}_\rho$~\cite{Slater1995,Slater1996}.
Just like $\mathbf{F}^{\g}$, this volume form is adjoint invariant under all $\G$ actions, meaning each $\M_{\rho}$ is a homogeneous, but not necessarily isotropic, space. 
D-optimal states can therefore be interpreted as members of the particular $\M_{\rho}$ which features with the largest quantum differential volume.
Notably when $\g$ contains \textit{all} observables over the Hilbert space and we consider the $\M_{\rho}$ of pure states, the eigenvalues of $\mathbf{F}^{\g}[\rho]$ are all equal to either a fixed value $c$ (related to our normalization on the $\{ \hat{G}_{\mu}\}$ or $0$. 
This leads to a homogeneous and isotropic quantum Jeffrey's prior, which closely resembles the Haar measure~\cite{Zoller2023,Mele2024,Song2024}.

\subsection{\texorpdfstring{$\SU{d}$ Expressions}{SU(d) Expressions}}
We can evaluate these two criteria explicitly for the case of $N$ many qudits, or $d$-level atoms, where the Lie algebra of interest is $\g = \su{d}$.
These are the generators of collective $d$-level rotations: $\hat{\mathcal{O}}_\mu = \sum_{j=1}^N \hat{\sigma}_\mu^{(j)}/2$ where $\mu=1,2,...d^2-1$ labels $\hat{\sigma}_\mu^{(j)}$, the $\mu^\mathrm{th}$ $\su{d}$ Gell-Mann operator acting on atom $j$~\cite{suppMatCons}.

Here, the highest weight sub-representation is the permutationally symmetric subspace, where $\zeta = \frac{1}{2} \frac{N(N+d)(d-1)}{d}$~\cite{SZOKE2019}.
When $N=1$, pure states have a fixed $\sum_\mu\langle\hat{G}_\mu \rangle^2 = 2(d-1)/d$~\cite{Toth3}. 
When $N\geq2$, however, the balanced GHZ state $\ket{\mathrm{GHZ}_\mathrm{Bal.}} = \frac{1}{\sqrt{d}} \sum_{n=1}^d \ket{n}^{\otimes N}$ has $\langle\hat{G}_\mu \rangle_\mathrm{Bal.} = 0$
for all $\mu$. 
We note that one may find many states satisfying $\langle \hat{G}_\mu \rangle_\mathrm{A\text{-}Opt} = 0$ for $\su{d}$.
Therefore,
\begin{equation}
\Tr_{\su{d}}(\mathbf{F}^{\su{d}}[\rho]) \leq  2 \frac{N(N+d)(d-1)}{d},
\end{equation}
with equality holding for $\rho_\mathrm{A\text{-}Opt}$.

We can use other bounds to derive an entanglement witness from the traced QFIM.
On this Lie algebra, $\mathbf{F}^{\su{d}}[\rho]$ bears witness to inter-qudit entanglement~\cite{Toth}, where if at least one eigenvalue has $\lambda_i \geq k N$ for $k$ dividing $N$, then $\rho$ is $k$-partite entangled~\cite{Hyllus}.
Per the argument about maximum weight representations, the equivalence class of generalized coherent spin states gives the upper bound for separable pure states.

These are states of the form $\ket{\mathcal{C}} = \ket{\psi}^{\otimes N}$ where $\ket{\psi}$ is any single particle state, and constitute their own equivalence class $\M_{\mathcal{C}}$ under $\SU{d}$ actions.
On $\M_\mathcal{C}$, $\mathbf{F}^{\su{d}}[\mathcal{C}]$ has $2(d-1)$ many eigenvalues of $N$, and $(d-1)^2$ many zero eigenvalues.
This is a result of the fact that $\su{d}$ has $d-1$ many unique $\su{2}$ sub-algebras, and so $\ket{\mathcal{C}}$ can evolve in $2$ orthogonal directions on any of the $d-1$ unique Bloch spheres~\cite{footnoteSubalgebras}.
Therefore, $\Tr_{\su{d}}(\mathbf{F}^{\su{d}}[\mathcal{C}]) = 2 N (d-1)$ and $\mathrm{pdet}(\mathbf{F}^{^{\su{d}}}[\mathcal{C}]) = N^{2(d-1)}$ for all coherent spin states.
This provides the entanglement witness
\begin{equation} \label{eq:TrAndDetCoherent}
\Tr_{\su{d}}(\mathbf{F}^{\su{d}}[\rho_\mathrm{separable}]) \leq 2 N(d-1),
\end{equation} 
where violation of this bound indicates \textit{at least} one of the eigenvalues scales $\lambda \geq N$~\cite{Pezze2009_HeisLim}.

One such set of entangled states are the N00N states~\cite{Combes2004}, or two-mode GHZ states~\cite{Zeilinger1998,Chen2010}, $\ket{\mathcal{N}} = ( \ket{\psi}^{\otimes N} + \ket{\perp}^{\otimes N} ) / \sqrt{2}$ where $\langle \perp | \psi \rangle = 0$. These states similarly form an equivalence class $\M_{\mathcal{N}}$ under $\SU{d}$ actions for $N\geq2$ (otherwise, $\M_{\mathcal{N}} = \M_{\mathcal{C}}$). On $\M_{\mathcal{N}}$, one eigenvalue of $\mathbf{F}^{\su{d}}[\mathcal{N}] $ is $N^2$, two are $N$, and $4 (d-2)$ many are $N/2$:
\begin{equation} \label{eq:TrNoon}
\Tr_{\su{d}}(\mathbf{F}^{\su{d}}[\mathcal{N}]) = N^2 + 2 N (d-1), \ 
\end{equation}
and $\mathrm{pdet}(\mathbf{F}^{\su{d}}[\mathcal{N}]) = N^4 \left(\frac{N}{2}\right)^{4(d-2)}$.
Interestingly, for $N=2$ and $d=3$ we see $\mathrm{pdet}(\mathbf{F}^{\su{d}}[\mathcal{C}]) = \mathrm{pdet}(\mathbf{F}^{\su{d}}[\mathcal{N}])$, indicating no analogous entanglement witness condition exists for the determinant.
We can compare these results to Ref.~\cite{Hyllus} for $d=2$ where the average QFI of separable qubit states is shown to be bounded by $2N(d-1)/\dim(\g) = 2N/3$, and entangled states are bounded the GHZ state; $N^2 + 2 N (d-2)/\dim(\g) = (N^2 + 2N)/3$.

\section{An Illustrative Example}
We now explore the consequences of our results in a physically motivated example. 
We consider $N$ permutationally symmetric three-level atoms, or qutrits, and compare two constructions of the QFIM from the Lie algebras $\g_1$ and $\g_2$.
First are the generators of the collective spin-1 dipole rotations, so $\g_1 = \su{2}$ is a reducible representation.
Second are the generators of arbitrary three-level rotations, so $\g_2=\su{3}$ is an irreducible representation.
Here, the three levels comprise the ground state manifold of an atom~\cite{Lyryl2025,PineiroOrioli,Silva,Yukawa,Smith2019}, but could also be states of polar molecules~\cite{Rey2021}, or recoil states of atomic momentum~\cite{Luo2024,Wilson2024}.

\begin{figure}
    \centering
    \includegraphics[width= 1 \columnwidth]{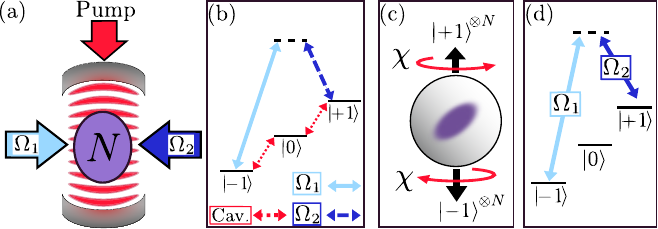}
    \caption{ (a) A cartoon of the example.
    (b) The atomic level structure. The virtual level (dotted line) is used for the two-photon interaction.
    The cavity is first pumped to seed the photon-mediated OAT, shown in (c) on the collective spin 1 pseudo-Bloch sphere. 
    Twisting occurs at rate $\chi$. 
    Second the atoms are directly driven by two lasers to drive a two photon transition at an effective rate $\Omega$, shown in (d) on the three hyperfine levels.
    }
    \label{fig:Schematic}
\end{figure}

The probe state is prepared by spin squeezing via a photon mediated interaction followed by a collective hyperfine transition via a two-photon transition, shown in~\cref{fig:Schematic} (a).
The hyperfine levels for atom $j$, and the collective generators of dipole rotations, are 
\begin{equation}
\ket{+1}_j,\  \ket{0}_j \  \mathrm{and} \  \ket{-1}_j, \quad \hat{S}_\mu = \sum_{j=1}^N \hat{s}_\mu^{(j)},
\end{equation}
where $\mu = x,y,z$ and $\hat{s}_\mu^{(j)}$ is the single atom operator: $\hat{s}_z^{(j)} = \ket{+1}_j\bra{+1}_j - \ket{-1}_j\bra{-1}_j$, $\hat{s}_x^{(j)} = ( \ket{+1}_j\bra{0}_j + \ket{0}_j\bra{-1}_j + \mathrm{H.c})/\sqrt{2}$, and $\hat{s}_y^{(j)} = -i [\hat{s}_z^{(j)}, \hat{s}_x^{(j)}]$.
The generators of hyperfine rotations, meanwhile, are
\begin{equation}
\begin{aligned}
\hat{Q}_\mu = \sum_{j=1}^N \frac{\hat{\sigma}_\mu^{(j)}}{2} \quad \mathrm{with} \quad \mu =1,...,8,
\end{aligned}
\end{equation}
with $\hat{\sigma}_\mu^{(j)}$ being the $\mu^\mathrm{th}$ $\su{3}$ Gell-Mann operator~\cite{Pfeifer2003} acting on atom $j$.
For example, $\hat{\sigma}^{(j)}_1 = \ket{+1}_j\bra{0}_j + \ket{0}_j\bra{+1}_j $. These operators are given explicitly in the SM~\cite{suppMatCons}.
This gives $\g_1 = \mathrm{span}( \{ \hat{S}_\mu \} ) \subset \g_2 = \mathrm{span}( \{ \hat{Q}_\mu\} )$.
 
The atomic structure is shown in~\cref{fig:Schematic} (b).
First, One-Axis Twisting (OAT) occurs at rate $\chi$ shown in ~\cref{fig:Schematic} (c), with the Hamiltonian $\hat{H}_S = \chi \hat{S}_z^2$. 
Subsequently, a two-photon transition is driven using virtual level, shown in~\cref{fig:Schematic} (d).
The effective transition rate is $\Omega$. 
This gives $\hat{H}_R = \Omega  \hat{Q}_3 $, where $\hat{Q}_3 = \sum_{j=1}^N ( \ket{+1}_j\bra{-1}_j + \ket{+1}_j\bra{-1}_j )/2$ is in $\g_2$. The full derivation of this model is found in the SM~\cite{suppMatCons}.
The atoms start in a coherent spin state in the $\hat{S}_x$-direction: $\ket{\psi_\mathrm{Init.}} = \exp[+i (\pi/2) \hat{S}_y] \ket{-1}^{\otimes N}$.
We calculate the QFIM for the state
$\ket{\alpha, \beta} \equiv \exp(-i \beta \hat{Q}_4 ) \exp(-i \alpha \hat{S}_z^2) \ket{\psi_\mathrm{Init.}}$
where $\alpha = \chi t_S $ and $\beta = \Omega t_R$.
As an aside, for spin-1 atoms the standard quantum limit, i.e., the QFI for collective dipole rotations of a coherent state, is $2N$, while the Heisenberg limit, i.e. the QFI for collective dipole rotations of a GHZ state~\cite{Pezze2009_HeisLim}, is $(2N)^2$.

\begin{figure*}
    \centerline{\includegraphics[width=\linewidth]{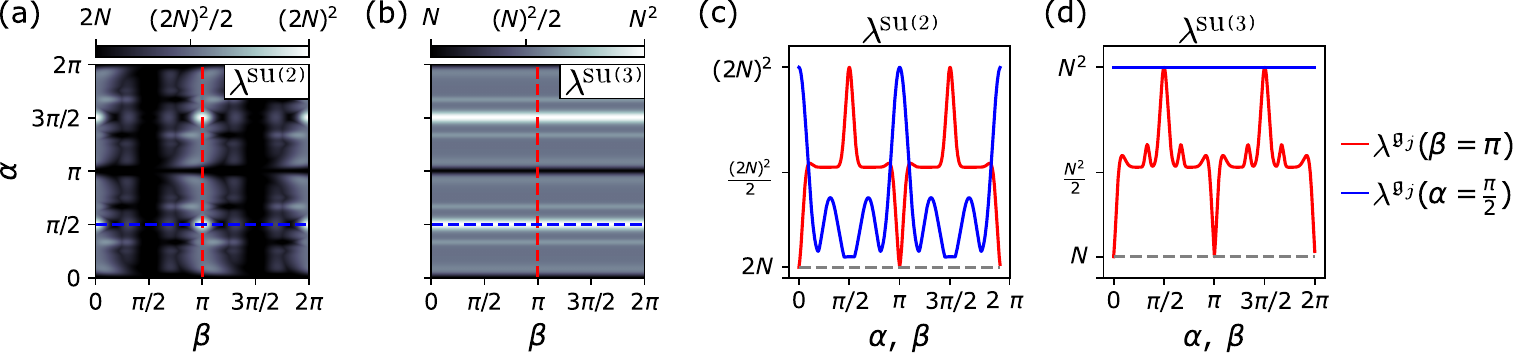}}
    \caption{
    (a) The largest eigenvalue of $\mathbf{F}^{\su{2}}[\rho]$ changes in response to both OAT and hyperfine rotations.
    (b) The largest eigenvalue of $\mathbf{F}^{\su{3}}[\rho]$ changes in response to OAT but is invariant to the hyperfine rotation, which is generated by $\g_2$.
    (c) Two slices of (a).
    For the slice along $\beta = \pi$ shows the expected QFI for the OAT. The slice along $\alpha=\pi/2$ shows that the $\g_2$ actions can modify the QFI with respect to $\g_1$ despite not creating nor destroying entanglement.
    (d) Two slices of (b). For the slice along $\beta = \pi$ the QFI is similar to that of $\g_1$, while for the slice along $\alpha=\pi/2$, the QFI is constant.
    }
    \label{fig:QFIMEig}
\end{figure*}

We numerically simulate these dynamics for $N = 10$ atoms and plot the maximum eigenvalue of the QFIM, $\lambda^{\g_j}$ in~\cref{fig:QFIMEig} for $\su{2}$ in (a), $\su{3}$ in (b).
There, $\lambda^{\su{2}}$ changes in response to both OAT (increasing $\alpha$) and the hyperfine rotation (increasing $\beta$).
However, $\lambda^{\su{3}}$ only changes in response to OAT and not $\hat{Q}_3 \in \su{3}$, according to the theorem.
Moreover, while the hyperfine rotation does not create interparticle entanglement it can obscure information from the $\su{2}$ QFIM.


This leads to a general observation: when $\g_1$ is a sub-algebra of $\g_2$, each equivalence class of $\g_2$ contains many equivalence classes of $\g_1$. Therefore $\g_2$ can modify the resources that $\mathbf{F}^{\g_1}[\rho]$ resolves.
This is apparent in~\cref{fig:QFIMEig} (c) and (d), which shows two slices of (a) and (b) for $\beta=\pi$ and $\alpha=\pi/2$.
The slice along $\beta=\pi$ shows the typical QFI for OAT~\cite{Pezze2018_NonclasAtoms,Wilson2023}.
The slice along $\alpha = \pi/2$ initially corresponds to a maximum of both $\lambda^{\su{2}}$ and $\lambda^{\su{3}}$ at $\beta=0$ at the respective Heisenberg limits. Then, as $\beta$ increases $\lambda^{\su{2}}$ decreases but $\lambda^{\su{3}}$ remains fixed.
This is a geometric rephrasing of results presented in Ref.~\cite{Knill2003}, where it is shown that entanglement should always be considered relative to a given Lie algebra of observables. 
The fact that  $\g_{1} \subset \g_{2}$ means that the $\g_{1}$ equivalence classes partition Hilbert space more finely than those of $\g_{2}$.
This is equivalent to the observation from~\cite{Knill2003} that when given two Lie algebras $\mathfrak{h}$ and $\g$ with $\mathfrak{h} \subset \g$, states are $\g$-coherent might not $\mathfrak{h}$-coherent, and therefore can be entangled with respect to $\mathfrak{h}$ but not $\g$.

Playing the same game as above, we can construct the algebra $\g_{3}$ which contains \textit{all} possible observables over $\mh$; here $D = \dim(\mh)$ and $\g_{3} = \su{D}$ is the defining representation. 
This is the set of \textit{all} possible unitary actions over $\mh$, and allows for the chain $\g_{1} \subset \g_{2} \subset \g_{3}$, where each $\g_{j}$ is a sub-algebra of each $\g_{j+1}$.
In this purely mathematical construction, it is ``as if'' the QFIM resolves one single $D$-level quDit where the equivalence classes are density matrices with the same spectrum, or von Nuemann entropy.
In this limit, there is no reservoir for meteorological resources except the purity of the state itself.
Thus, $\Tr_{\g_{3}}(\mathbf{F}^{\g_3}[\rho]) \leq 2(D-1)$, with equality only for pure states, because all pure states are $\g_3$-coherent states.
The example only considers unitary dynamics on the permutationally symmetric subspace~\cite{Mathur2010} for $N=10$. 
Therefore, $D=(N+1)(N+2)/2=66$, and $\Tr_{\g_3}(\mathbf{F}^{\g_3}[\rho]) = 2(D-1) = 130$ is fixed for all dynamics because OAT belongs to $\g_3$.

\begin{figure*}
    \centerline{\includegraphics[width=\linewidth]{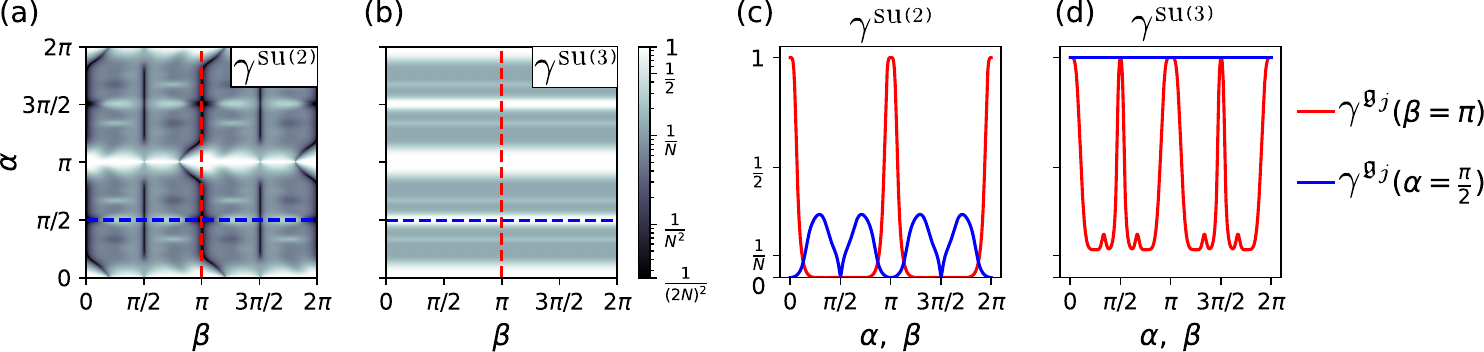}}
    \caption{(a) $\gamma^{\su{2}}(\alpha,\beta)$, the metrological incompatibility of dipole rotations, in a log-scale with a cutoff of $1/(2N)^2$. 
    The incompatibility changes in response to both OAT and hyperfine rotations.
    (b) $\gamma^{\su{3}}(\alpha,\beta)$, the metrological incompatibility of hyperfine rotations in the same log scale, which is now invariant to hyperfine rotations.
    (c) The same two slices as~\cref{fig:QFIMEig}(c) and (d), where the metrological incompatibility drops fully to zero on the plateau of OAT---therefore one could simultaneously estimate all three vector directions of a magnetic field with no measurement induced back action.
    (d) The same two slices for the incompatibility of $\g_2$. }
    \label{fig:UCT}
\end{figure*}

In ~\cref{fig:UCT}, we plot the metrological incompatibility parameter~\cref{eq:Incompatibility} for $\su{2}$ and $\su{3}$ in~\cref{fig:UCT}.
There, we see that the same invariance as found in the QFIM is apparent.
OAT introduces entanglement which modifies the metrological incompatibility of both the dipole rotations, $\g_1$, and hyperfine rotations, $\g_2$ while the subsequent hyperfine rotation only affects the incompatibility of $\g_1$.
Here, we also see that the metrological incompatibility can be made to vanish entirely on $\g_1$ under OAT, but not on $\g_2$.
One may consider larger and larger Lie algebras where each contains the last, such as the chain $\g_1 \subset \g_2 \subset \g_3$.
This will always give access to operators that might be ``more optimal'' under the QFIM, however the metrological incompatibility necessarily also increases under this chain.

The incompatibility begins to decrease on the same timescale as the optimal squeezing time, meaning one might explore methods to such as Floquet driving to create such a state even in the presence of decoherence~\cite{Reilly2024}. 
The goal of minimizing the incompatibility fits in with the geometric criteria in~\cref{tab:Criteria}, but has no classical analog.
For sake of consistency, one might call it Q-optimality, where geometrically it corresponds to a vanishing gauge-curvature~\cite{Carollo}.

When the Uhlmann curvature $\mathbf{U}^{\g}[\rho]$ vanishes, it vanishes on the entire equivalence class $\M_\rho$.
Therefore, any state $\rho$ with vanishing $\mathbf{U}^{\g}[\rho]$ immediately defines the manifold $\mathcal{M}_{\rho}$ which comprises a pseudo Quantum Mechanics Free Subspace (QMFS) for $\g$, where every parameter can be simultaneously optimally estimated.
We specify \textit{pseudo}-QMFS because it is the parameter estimation which evades measurement-induced back action rather than the observables themselves~\cite{Caves2012}.
This is quantified by the CFIM for a Positive Operator Valued Measure (PVOM), $\{ \hat{\Pi} \}$, where
\begin{equation}
\mathcal{I}[\rho|\{\hat{\Pi}\}]_{\mu\nu} = -\sum_{m} \frac{ \Tr_{\mh}( [\hat{G}_\mu, \rho] \hat{\Pi}_m ) \Tr_{\mh}( [\hat{G}_\nu, \rho] \hat{\Pi}_m )}{\Tr_{\mh}( \rho \hat{\Pi}_m )}.
\end{equation}
The CFIM provides the asymptotic bound the precision of multiparameter estimation~\cite{Liu2019} and, when $\mathbf{U}^{\g}[\rho]$ vanishes, there exists an optimal POVM under which the CFIM is equal to the QFIM.
In principle the optimal POVM may differ for each $\rho$.
Nonetheless, for each $\rho \in \M_\rho$ there exists $\{\hat{\Pi}_\mathrm{Opt}\}_\rho$ with $\mathcal{I}[\rho|\{\hat{\Pi}_\mathrm{Opt}\}_\rho]_{\mu\nu} = \mathcal{F}^{\g}[\rho]_{\mu\nu}$.

%

\section{Metrological Liouville's Theorem}
When these results are all taken together, we find a metrological analog to Liouville's theorem~\cite{Liouville1838,Song2024}:
$\M_\rho$ is the phase space of all evolution unitarily generated by $\g$ starting from $\rho$.
When the QFIM and UCT are constructed from $\g$, $\M_\rho$ comprises a manifold whereupon the Lie group, $\G$ does not compress volume, given by $\mathrm{pdet}(\mathbf{F}^\g[\rho])$, nor stretch distances, given by the spectrum of $\mathbf{F}^\g[\rho]$, nor curve space, given by $\mathbf{U}^{\g}[\rho]$.
These geometric quantities are invariant, and this invariance corresponds to the manifold symmetries of $\M_\rho$ itself.
In the ad-nauseam limit that one includes every observable over Hilbert space, this invariance principle extends to all of unitary mechanics.

\section{Conclusion and outlook}
We have proven a simple yet previously unexplored conservation law of the QFIM, and connected it to the underlying geometrical symmetries of quantum state space.
Furthermore, through these symmetries the UCT inherits this same invariance.
This conservation law arises when one constructs the SLDs which are used to calculate the QFIM and UCT from an underlying Hermitian Lie algebra.
This Hermitian Lie algebra naturally foliates the space of density matrices into equivalence class manifolds.
On these manifolds, a metrological analog to Liouville's theorem plays out where distances, curvatures, and volumes as diagnosed by the QFIM and UCT are incompressible.

From this invariance, we identified the geometric descriptions of quantum analogs to the classical optimality criteria typically given by functionals of the CFIM~\cite{kiefer1959}.
Using this, we derived conditions based on the trace of the QFIM to formulate a common entanglement witness and connected the determinant of the QFIM to the quantum Jeffrey's prior on quantum state manifolds.
We showed that in the absurd limit for which one constructs the QFIM from all possible observables in Hilbert space, one recovers a unitarily invariant volume form that is reminiscent of the Haar measure.

We explored the consequences of this in the context of polar spin-squeezing where we showed the interplay of three Lie algebras, each one containing the last, and each generating a different partition of the space of density matrices.
There we showed that the incompatibility can actually vanish in an (idealized) physical system, thereby leading to a pseudo-QMFS where one could simultaneously estimate every parameter for a given $\rho$.

The QFIM and UCT are the real and imaginary components of the Quantum Geometric Tensor (QGT), a core object to quantum state geometry.
Because both the QFIM and UCT are invariant, the QGT is as well.
We focused on the QFIM and UCT objects separately as means for quantum state characterization and deriving useful intuition, but an interesting question is the implications of the invariant QGT as a whole.
There might be opportunities to leverage this invariance in situ with quantum learning techniques~\cite{Beckey2022,Zhang2025} to greatly simplify the characterization of quantum states.
One could explore this invariance in experimentally accessible quantities like the sub-quantum Fisher information~\cite{Beckey2021} or the semi-classical QGT~\cite{Pezze2025}.
Of particular interest in extending these results is the apparent connections between the Haar-measure, an important tool in quantum computing theory~\cite{Mele2024}. The existence of conserved quantities within each equivalence class also may open the door to novel subspace based quantum computing benchmarking methods \cite{Baldwin2020}.

Lastly, we presented these results in the context of finite dimensional Hilbert spaces.
They do not extend straightforwardly to infinite dimensional Hilbert spaces because one loses the straightforward orthogonality condition provided by the Hilbert-Schmidt inner product and instead must use the Killing form on the Lie algebra.
In Ref.~\cite{Wilson2025} we develop the mathematical formalism for \textit{any} Riemannian metric and extend an analogous conservation law to the infinite dimensional case, where these same invariance principles are non-trivially modified.

\,\newline
\noindent {\em Acknowledgments:} 
The authors would like to thank Shawn Geller, Maria Luisa Chiofalo, Augusto Smerzi, Athreya Shankar, Lyryl Vaecairn and Jarrod T. Reilly for useful discussions.
This work was supported by the National Science Foundation under QLCI Award No. OMA 2016244 and Grant No. PHY 2317149. Additional funding is provided by the US
Department of Energy, Office of Science, National Quantum Information Science Research Centers, Quantum
Systems Accelerator. 

\input{main.bbl}
\vspace{2em}

\newpage\newpage

\appendix

In appendix~\ref{sec:InnerProd} we outline the trace-inner product and describe the normalization, $C$, from the main text.
In appendix~\ref{sec:TraceandDet}, we show the explicit calculation of the trace and determinant for coherent and Noon states, stated in the main text.
In appendix~\ref{sec:AvgRes}, we directly calculate the average response of the quantum state.
In appendix~\ref{sec:MainOps}, we provide a formal construction of the operator basis of $\su{d}$, $\g_{1}$, $\g_{2}$, and $\g_{3}$, used in the main text.
In appendix~\ref{sec:Model}, we provide a detailed derivation of the model used to explore the dynamics presented in the main text.

\section{The Hilbert Schmidt Inner Product}
\label{sec:InnerProd}
Throughout the main text, we make use of the fact that one can fix the normalization of operators using the Hilbert Schmidt inner product:
\begin{equation}
\langle \hat{A}, \hat{B} \rangle_{\mathrm{HS}} \equiv \Tr_{\mh}( \hat{A}^\dagger \hat{B} ) .
\end{equation}
In the main text, we only consider this as an inner product with Hermitian operators, and so we drop the complex conjugate on the first term.
We use this to define the inner product for orthonormal basis elements of a Lie algebra:
\begin{equation}
\Tr_{\mh}( \hat{G}_\mu \hat{G}_\nu ) = C \delta_{\mu\nu}
\end{equation}
where $C \in \mathbb{R}$ and $\delta_{\mu\nu}$ is the delta function.

As mentioned at in the paper, we use this inner product to raise an index on the QFIM in order to make it a well-defined inner product.
This means we define
\begin{equation} \label{eq:HS_Metric}
g_{\mu\nu} \equiv \Tr_{\mh}( \hat{G}_\mu \hat{G}_\nu ),
\end{equation}
whereupon
\begin{equation}
F^{\g}[\rho]^\mu_\nu = C \ g^{\mu\alpha} \ \F^\g[\rho]_{\alpha\nu} 
\end{equation}
provides the matrix elements of the matrix $\mathbf{F}^{\g}[\rho]$.
Here, $g^{\mu\alpha}$ is the inverse of~\cref{eq:HS_Metric} such that
\begin{equation}
g^{\mu\alpha} g_{\alpha\nu} = \delta^\mu_\nu,
\end{equation}
for the delta function $\delta^\mu_\nu$.
This means that
\begin{equation}
g^{\mu\alpha} = \frac{1}{C} \delta^{\alpha \mu}
\end{equation}
in the basis we chose.
As a result
\begin{equation}
\begin{aligned}
F^{\g}[\rho]^\mu_\nu =& C \ g^{\mu\alpha} \ \F^\g[\rho]_{\alpha\nu} \\
=&  \delta^{\mu\alpha} \ \F^\g[\rho]_{\alpha\nu}
\end{aligned} 
\end{equation}
which is exactly what we want, so that the eigenvalues match the Cr\'amer-Rao Bound.

\section{Construction of the various Lie Algebras}
\label{sec:MainOps}

Throughout the main text, we consider some operators which close to form a Lie algebra. We only consider Hermitian Lie algebra's.
This is a set of operators, $\g$, that are Hermitian: $\hat{A} = \hat{A}^\dagger$ for $\hat{A} \in \g$, and with the following properties:
\begin{equation}
\begin{aligned}
\hat{A}, \hat{B} \in \g \quad \mathrm{implies} \quad [\hat{A},\hat{B}] = i \hat{H} \quad \mathrm{with} \quad \hat{H} \in \g .
\end{aligned}
\end{equation}
When we choose the linearly independent and normalized basis for a Lie algebra, we mean the following:
\begin{equation}
\begin{aligned}
\mathrm{for\ any} \ \hat{A} \in \g, \quad
\hat{A} = \sum_{\mu} \frac{1}{C} \Tr_{\mh}(\hat{G}_\mu \hat{A}) \hat{G}_\mu, \\
\mathrm{such \ that} \quad \Tr_{\mh}(\hat{G}_\mu \hat{G}_\nu) = C \delta_{\mu\nu}, \ \mathrm{for} \ C\in\mathbb{R},
\end{aligned}
\end{equation}
and therefore $\mathrm{span}( \{\hat{G}_\mu\} ) = \g$.
We can now outline the constructions of the basis elements we use in the main text.

\subsection{\texorpdfstring{$\g = \su{d}$}{g = su(d)}}\label{sec:su(d)}

Now, we will construct the Lie algebra of the general $N$ particle representation of $\su{d}$.
Here, we will construct the $N$ particle representation of these operators according to
\begin{equation}
\hat{\mathcal{O}}_\mu \equiv \sum_{j=1}^N \frac{\hat{\sigma}^{(j)}_\mu}{2}
\end{equation}
where $\hat{\sigma}^{(j)}_\mu$ is, once again, the $d\times d$ Gell-Mann operator for particle $j$.
Here, we label the individual atom states by the indices $m$ and $n$ that range from $1$ to $d$, where $m<n$.
Here we will construct the single particle operators as matrices, but we recommend Ref.~\cite{Pfeifer2003} for a more complete description.
In this construction, we'll drop the index $j$ labeling which particle they act on, for conciseness.
To find a basis of $\su{d}$, for which there will be $d^2 - 1$ many operators, we will first construct the ``x''-like and ``y''-like operators.
We do this construction pairwise for the single particle states, labeled $(n,m)$, of which there are $d(d-1)/2$ many pairs.
We can index each pair of states by number $k(m,n)$, where
\begin{equation} \label{eq:k_index}
k(m,n) = \frac{(n-1)(n-2)}{2} + m
\end{equation}
where $1 \leq m < n \leq d$.
Furthermore, each pair of states has an operator
\begin{equation}
\hat{E}_{nm} = \op{n}{m}
\end{equation}
whereupon we use the index, $k(n,m)$ from~\cref{eq:k_index} to label
\begin{equation}
\hat{\sigma}_{2k-1} = (\hat{E}_{nm} + \hat{E}_{mn}), \hat{\sigma}_{2k} = -i(\hat{E}_{nm} - \hat{E}_{mn}),
\end{equation}
where we've suppressed the dependence on $n$ and $m$ in $k$.
Now, $k$ runs from $1$ to $d(d-1)/2$ labeling each pair, giving $d(d-1)$ many operators.
This leaves the $d-1$ many diagonal operators, which we construct the following way:
\begin{equation}
  \hat{\sigma}_{\ell} = \sqrt{\frac{2}{\ell(\ell+1)}}\Bigl(\hat{E}_{11} +\hat{E}_{22}+\cdots+\hat{E}_{\ell\ell}- \ell \hat{E}_{\ell+1,\ell+1} \Bigr),
\end{equation}
where $\ell$ runs from $d(d-1) + 1$ to $d^2-1$.
In the case that $d=3$, this matches~\cref{sec:su3}.

It is often not feasible to simulate the whole of these dynamics, which grow as $N^d$.
Instead, we can also do this construction for just the symmetric subspace--which is what we simulated in the main text.
For this construction, we use the list of annihilation (creation) operators $\vec{a}$ ($\vec{a}^\dagger$), where $\hat{a}_m$ ($\hat{a}^\dagger_m$) annihilates (creates) a particle in the $m^\mathrm{th}$ state. 
When $d=3$, $\hat{a}_1 = \hat{a}$, $\hat{a}_2 = \hat{b}$, $\hat{a}_3 = \hat{c}$ above.
We build the operators in the same manner.
Let $\boldsymbol{\sigma}_\mu$ be the matrix representation of the operator $\hat{\sigma}_\mu$, then simply
\begin{equation}
\hat{\mathcal{O}}_\mu \equiv \vec{a}^\dagger \boldsymbol{\sigma}_\mu \vec{a} / 2
\end{equation}
and the normalization becomes
\begin{equation}
\begin{aligned}
\Tr_{\mh_\mathrm{sym}}( \hat{\mathcal{O}}_\mu \hat{\mathcal{O}}_\nu ) = \frac{\delta_{\mu\nu}}{2} \begin{pmatrix}
N+d \\
d+1
\end{pmatrix} = \frac{\delta_{\mu\nu}}{2(d+1)!} \prod_{j=0}^d(N+j)
\end{aligned}
\end{equation}
where $\begin{pmatrix}
a \\
b
\end{pmatrix} = \frac{a!}{(a-b)! b!}$ is the binomial coefficient, and we can identify $C = \frac{1}{2} \begin{pmatrix}
N+d \\
d+1
\end{pmatrix}$.
We note that this normalization does not match the normalization of that given in Ref.~\cite{Wilson2023}, Eq. (S47) of the supplement where there was a typo, missing the factor of $1/2$ and with the wrong starting index on the product.

\subsection{\texorpdfstring{$\g_{1} = \su{2}$}{g1 = su(2)}}\label{sec:su2}
Here, we explicitly construct the dipole operators considered in the example of the main text: $\g_1 = \su{2}$, or the operators which generate collective spin-1 rotations. 
The representation of this algebra is spanned by three operators, $\hat{S}_{x}$, $\hat{S}_{y}$ and $ \hat{S}_{z}$. 
These are give by
\begin{equation}
\begin{aligned}
\hat{S}_\mu \equiv \sum_{j=1}^N \hat{s}_\mu^{(j)} \quad \mathrm{for} \quad \mu = x,y,z
\end{aligned}
\end{equation}
where
\begin{equation}
\begin{aligned}
\hat{s}_x^{(j)} \equiv& ( \ket{+1}_j\bra{0}_j + \ket{0}_j\bra{-1}_j + \mathrm{H.c})/\sqrt{2}, \\
\hat{s}_y^{(j)} \equiv& -i ( \ket{+1}_j\bra{0}_j + \ket{0}_j\bra{-1}_j - \mathrm{H.c})/\sqrt{2}, \\
\hat{s}_z^{(j)} \equiv& \ket{+1}_j\bra{+1}_j - \ket{-1}_j\bra{-1}_j. \\
\end{aligned}
\end{equation}
For the purpose of numerical simulation, we make use of Schwinger bosons following the procedure of Ref.~\cite{Mathur2010}.
These are operators which create (or annihilate) atoms such that
\begin{equation}
\begin{aligned}
\hat{a}^\dagger \ket{\mathrm{vac}} = \ket{+1}\quad \hat{b}^\dagger \ket{\mathrm{vac}} = \ket{0}\quad \hat{c}^\dagger \ket{\mathrm{vac}} = \ket{-1}.\\
\end{aligned}
\end{equation}
Utilizing these operators we explicitly have
\begin{equation}
    \begin{aligned}
    \hat{S}_{x} &= \frac{1}{\sqrt{2}}( \hat{a}^\dagger \hat{b} + \hat{b}^\dagger \hat{c}  + \hat{b}^{\dagger} \hat{a} + \hat{c}^{\dagger}\hat{b}) \\ 
    \hat{S}_{y} &= \frac{1}{i \sqrt{2}}( \hat{a}^\dagger \hat{b} + \hat{b}^\dagger \hat{c}  - \hat{b}^{\dagger} \hat{a} - \hat{c}^{\dagger}\hat{b}) \\ 
    \hat{S}_{z} &= \hat{a}^{\dagger}\hat{a} - \hat{c}^{\dagger}\hat{c}
    \end{aligned}
\end{equation}
These operators satisfy the usual commutation relations $[\hat{S}_x,\hat{S}_y] = i \hat{S}_z$.
This can be re-written tensorially
$[\hat{S}_{i},\hat{S}_{j}] = i \epsilon_{i j}^{k}\hat{S}_{k}$, with $\epsilon_{i j}^{k}$ the totally antisymmetric tensor. 
This is an orthonormal set of operators on $\mh$: 
\begin{equation}
    \begin{aligned}
\mathrm{Tr}_{\mh_\mathrm{sym}}(\hat{J}_{\mu}&\hat{J}_{\nu}) = \\
& \left(\frac{N(N+1)(N+2)(N+3)}{12}\right) \delta_{\mu \nu},
    \end{aligned}
\end{equation}
where we can identify $C \equiv N(N+1)(N+2)(N+3)/12 $.

\subsection{\texorpdfstring{$\g_{2} = \su{3}$}{g2 = su(3)}}\label{sec:su3}
The second algebra are the collective hyperfine transitions for the three levels, corresponding to $\g_{2} = \su{3}$. 
To build the $\su{3}$ algebra, we can once again use sums over over the single particle operators:
\begin{equation}
\begin{aligned}
\hat{Q}_\mu \equiv \sum_{j=1}^N \hat{\sigma}^{(j)}_\mu / 2,
\end{aligned}
\end{equation}
where we use $\sigma$ here to evoke the similarities to the Pauli-matrices and $Q$ because these are like quadrapole operators. Above, we used $s$ because they are the spin-1 Pauli-matrices.
Here, $\hat{\sigma}^{(j)}_\mu$ is the $\mu^\mathrm{th}$ Gell-Mann matrix.
Explicitly, in matrix form, these are
\begin{equation}
\begin{aligned}
\boldsymbol{\sigma}_1 &= \begin{pmatrix}
0 & 1 & 0 \\
1 & 0 & 0 \\
0 & 0 & 0
\end{pmatrix}, &
\boldsymbol{\sigma}_2 &= \begin{pmatrix}
0 & -i & 0 \\
i & 0 & 0 \\
0 & 0 & 0
\end{pmatrix}, \\
\boldsymbol{\sigma}_3 &= \begin{pmatrix}
0 & 0 & 1 \\
0 & 0 & 0 \\
1 & 0 & 0
\end{pmatrix}, &
\boldsymbol{\sigma}_4 &= \begin{pmatrix}
0 & 0 & -i \\
0 & 0 & 0 \\
i & 0 & 0
\end{pmatrix}, \\
\boldsymbol{\sigma}_5 &= \begin{pmatrix}
0 & 0 & 0 \\
0 & 0 & 1 \\
0 & 1 & 0
\end{pmatrix}, &
\boldsymbol{\sigma}_6 &= \begin{pmatrix}
0 & 0 & 0 \\
0 & 0 & -i \\
0 & i & 0
\end{pmatrix}, \\
\boldsymbol{\sigma}_7 &= \begin{pmatrix}
1 & 0 & 0 \\
0 & -1 & 0 \\
0 & 0 & 0
\end{pmatrix}, &
\boldsymbol{\sigma}_8 &= \frac{1}{\sqrt{3}} \begin{pmatrix}
1 & 0 & 0 \\
0 & 1 & 0 \\
0 & 0 & -2
\end{pmatrix}
\end{aligned}
\end{equation}
where we've labeled them by ``x-y'' like pairs first, then the operators that act like a ``z`` operator--i.e. the diagonal operators.
For posterity, we would have that $\hat{\sigma}^{(j)}_7 = \ket{+1}_j\bra{+1}_j - \ket{0}_j\bra{0}_j$.
We can use this matrix form, the creation (annihilation) operators from the previous section, and the vectors $\vec{a} \equiv ( \hat{a}, \hat{b}, \hat{c} )^T$, $\vec{a}^\dagger \equiv ( \hat{a}^\dagger, \hat{b}^\dagger, \hat{c}^\dagger )$ to define any $\hat{Q}_\mu$ on the permutationally symmetric representation of $N$ many three-level atoms as $\hat{Q}_\mu = \frac{1}{2}\vec{a}^\dagger \cdot \boldsymbol{\sigma}_\mu \cdot \vec{a}$.
This gives
\begin{equation}
\begin{aligned}
    \hat{Q}_1 &= \frac{1}{2} (a^\dagger b + b^\dagger a), & \hat{Q}_2 &= \frac{1}{2i} (a^\dagger b - b^\dagger a), \\
    \hat{Q}_3 &= \frac{1}{2} (a^\dagger c + c^\dagger a), & \hat{Q}_4 &= \frac{1}{2i} (a^\dagger c - c^\dagger a), \\
    \hat{Q}_5 &= \frac{1}{2} (b^\dagger c + c^\dagger b), & \hat{Q}_6 &= \frac{1}{2i} (b^\dagger c - c^\dagger b), \\
    \hat{Q}_7 &= \frac{1}{2} (a^\dagger a - b^\dagger b), & \hat{Q}_8 &= \frac{1}{2\sqrt{3}} (a^\dagger a + b^\dagger b - 2c^\dagger c).
\end{aligned}
\end{equation}
We note that $\g_1 \subset \g_2$ where
\begin{equation}
\begin{aligned}
\hat{S}_x =& \sqrt{2}(\hat{Q}_1 + \hat{Q}_5), \\
\hat{S}_y =& \sqrt{2}(\hat{Q}_2 + \hat{Q}_6), \\
\hat{S}_z =& \hat{Q}_7 + \sqrt{3} \hat{Q}_8.
\end{aligned}
\end{equation}. 
These operators have the same trace norm:
\begin{equation}
\begin{aligned}
\mathrm{Tr}_{\mh_{sym}}(Q_{\mu}Q_{\nu}) = \left(\frac{N(N+1)(N+2)(N+3)}{48}\right) \delta_{\mu \nu},
\end{aligned}
\end{equation}
whereupon we can identify $C = \frac{N(N+1)(N+2)(N+3)}{48}$.
The fact that $\Tr_{\mh_\mathrm{sym}}(\hat{S}_x^2) = 4 \Tr_{\mh_\mathrm{sym}}(\hat{Q}_1^2)$ corresponds to the fact that the standard quantum limit and Heisenberg limit for $\g_1$ are $2N$ and $4 N^2$ respectively, whereas they are $N$ and $N^2$ respectively for $\g_2$.
The norm used for $\g_2$ is the same norm outlined in Ref.~\cite{Wilson2023} and is, in some sense, canonical by fixing the standard quantum limit and Heisenberg limit at $N$ and $N^2$.

\subsection{\texorpdfstring{$\g_{3} = \uu{\mh_\mathrm{sym}}$}{g3 = u(Hsym)}}\label{sec:u(hysm)}
Now we consider the third Lie algebra from the example in the main text. 
This is the set of all possible observables on the symmetric subspace, denoted $\g_3 = \uu{\mh_\mathrm{sym}}$ where no representation is needed. 
In the limit that $d=D$ and $N=1$ this matches~\cref{sec:su(d)}, although there the notion of ``particle number'' being $N=1$ is a purely mathematical construction.

This is the algebra containing every Hermitian observable in $\mh_\mathrm{sym}$, of which there are $D^2$ including the identity.
We identify this basis by the operators $\{\hat{T}_\mu\}$.
First, we remove the identity and give it the label $\hat{T}_0$, so we are actually after $\su{\mh_\mathrm{sym}}$, the set of all possible traceless Hermitian observables.
The size of the remaining set is $\dim(\su{\mh_\mathrm{sym}})  = (\mathrm{dim}(\mh_\mathrm{sym})^{2} - 1)$, where $\mathrm{dim}(\mh_\mathrm{sym})= (N+1)(N+2)/6$ for $N$ many three level atoms, meaning that unlike our other two algebras, the number of basis elements here depends on our number of atoms.
We first number all the basis states of $\mh_\mathrm{sym}$ by $\ket{j}$ with $j$ from $j=1$ to $j=\mathrm{dim}(\mh) \equiv D$.
This numbering is arbitrary, but then we subsequently build operators on these states following the construction of~\cref{sec:su(d)}. 

These operators are traceless and mutually orthonormal under the trace inner product.
That is to say:
\begin{equation}
\Tr_{\mh_\mathrm{sym}}( \hat{T}_\mu \hat{T}_\nu ) = \frac{\delta_{\mu\nu}}{2},
\end{equation}
whereupon $C=2$.
Clearly, both $\g_1$ and $\g_2$ are sub-algebras of $\g_3$.

\section{Trace and Determinant Expressions}
\label{sec:TraceandDet}
In this section, we derive the expressions we gave for the trace and determinant of the coherent spin state, and of the Noon state in the main text.
We also derive the condition for the entanglement witness using the trace.
Throughout, we take $\g$ to be the $N$ particle representation of $\su{d}$, given in~\cref{sec:su(d)}.
We also use
\begin{equation}
\mathcal{F}[\rho]_{\mu\nu} = 2 \langle \{\hat{G}_\mu,\hat{G}_\nu \rangle_\rho - 4 \langle \hat{G}_\mu \rangle_\rho \langle \hat{G}_\nu \rangle_\rho
\end{equation}
for pure states.

\subsection{The Coherent Spin State}
A generalized Coherent Spin State (CSS) is given by $\ket{\mathcal{C}} = \ket{\psi}^{\otimes N}$ where $\ket{\psi}$ is any single particle pure state~\cite{Nemoto2000}.
We want to classify the trace and determinant of the QFIM of this state, to do so we will use the observation made in Ref.~\cite{Knill2003}.
Namely, every $\ket{\mathcal{C}}$ is separable and therefore it is the unique groundstate of the sum of single particle operators.
This is a long winded way to say, we might as well label our single particle states such that $\ket{\psi} = \ket{1}$, and so $\ket{\mathcal{C}} = \ket{1}^{\otimes N}$.
This let's us turn the calculation of the trace and determinant into a counting problem.

First, we note that $\ket{\mathcal{C}}$ is an eigenstate of $d-1$ many diagonal operators in this basis.
Second, we note that this state is in the kernel of $(d-1)(d-2)$ many non-diagonal operators--i.e. the x and y-like operators without an element acting on $\ket{1}$ but acting on the $d-1$ many remaining states.
This leaves us with $2(d-1)$ many remaining operators in the basis of the form
\begin{equation}
\begin{aligned}
\hat{A}^{(m)}_x &= \sum_{j=1}^N \frac{1}{2} \left( \ket{1}_j\bra{m}_j + \mathrm{H.c.} \right) \\
\hat{A}^{(m)}_y &= \sum_{j=1}^N \frac{1}{2i} \left( \ket{1}_j\bra{m}_j - \mathrm{H.c.} \right)
\end{aligned}
\end{equation}
for $m=2,d$.
We also note that the offdiagonal QFIM between any of these operators is zero. This is to say:
\begin{equation}
\begin{aligned} \label{eq:NoOffDiagonals}
4\left( \left\langle \frac{\hat{A}^{(m)}_{\mu} \hat{A}^{(n)}_{\nu} + \hat{A}^{(n)}_{\nu} \hat{A}^{(m)}_{\mu}}{2} \right\rangle_{\mathcal{C}} - \langle \hat{A}^{(m)}_{\mu} \rangle_{\mathcal{C}} \langle \hat{A}^{(n)}_{\nu} \rangle_{\mathcal{C}} \right) = 0
\end{aligned}
\end{equation}
for $\mu,\nu = x, y$ and $n,m=2,d$ but either $n\neq m$ or $\mu\neq\nu$ (i.e. $\hat{A}^{(n)}_{\nu} \neq \hat{A}^{(m)}_{\mu}$).
For which $\ket{\mathcal{C}}$ has a QFI of $N$.
This yields $2(d-1)$ many eigenvalues of the QFIM of $N$ for a coherent spin state. 
Therefore:
\begin{equation}
\begin{aligned}
\Tr_{\su{d}}(\mathbf{F}^{\su{d}}[\mathcal{C}]) =& 2 N (d-1),
\\
\mathrm{pdet}(\mathbf{F}^{\su{d}}[\mathcal{C}]) =& N^{2(d-1)}.
\end{aligned}
\end{equation} 

An alternative argument goes as follows.
$\ket{C}$ is the (non-zero) eigenstate of $d-1$ many operators--which is the dimension of the maximal Cartan sub-algebra.
The Lie algebra $\su{d}$, for the same reason, has $d-1$ many unique $\su{2}$ sub-algebras.
This means $\ket{C}$ sits at the ``North pole'' of $d-1$ many collective Bloch spheres.
This gives $2(d-1)$ many unique directions for $\ket{C}$ to evolve in.
Furthermore, if we normalize everything properly, it will evolve in each direction at a rate $N$ corresponding to a QFI of $N$.
Notably, for multiparameter estimation, $d-1$ of these QFI's could be simultaneously estimated with commuting measurements.

\subsection{The Noon State}
In the text, we give the definition for the generalzied Noon state, which we construct as the sum of two orthogonal CSS:
\begin{equation}
\begin{aligned}
\ket{\mathcal{N}} = ( \ket{\psi}^{\otimes N} + \ket{\perp}^{\otimes N} ) / \sqrt{2}
\end{aligned}
\end{equation}
where $\ket{\psi}$ is a single particle state, just like in $\ket{\mathcal{C}}$, and $\ket{\perp}$ is any perpendicular single particle state to $\ket{\psi}$.
We can use the same labeling argument as above to identify the state $\ket{\psi} = \ket{1}$ and $\ket{\perp} = \ket{2}$.
These states are different eigenstates of the same $d-1$ operators.
From the construction in~\cref{sec:su(d)}, they are degenerate eigenstates of $d-2$ of these, and therefore $\ket{\mathcal{N}}$ is an eigenstate of these $d-2$ operators as well.
$\ket{1}^{\otimes N}$ and $\ket{2}^{\otimes N}$ are not, however, degenerate eigenstates of the operator:
\begin{equation}
\hat{B}_z \equiv \sum_{j=1}^N \frac{1}{2} \left( \ket{1}_j\bra{1}_j - \ket{2}_j\bra{2}_j \right),
\end{equation}
of which they are the optimal state for sensing a single parameter differential shift, which is to say they have a QFI of $N^2$.
Furthermore, they have a QFI of $N$ with respect to two more operators:
\begin{equation}
\begin{aligned}
\hat{B}_x &= \sum_{j=1}^N \frac{1}{2} \left( \ket{1}_j\bra{2}_j + \mathrm{H.c.} \right) \\
\hat{B}_y &= \sum_{j=1}^N \frac{1}{2i} \left( \ket{1}_j\bra{2}_j - \mathrm{H.c.} \right).
\end{aligned}
\end{equation}
Now, there are $d^2-4$ many more elements to consider in the construction of the QFIM.
First, for $(d-2)(d-3)$ of these, $\ket{\mathcal{N}}$ is in the kernel.
This leaves $4(d-2)$ many remaining operators.
These operators are of the fallowing form:
\begin{equation}
\begin{aligned}
\hat{A}^{(n,m)}_x &= \sum_{j=1}^N \frac{1}{2} \left( \ket{n}_j\bra{m}_j + \mathrm{H.c.} \right) \\
\hat{A}^{(n,m)}_y &= \sum_{j=1}^N \frac{1}{2i} \left( \ket{n}_j\bra{m}_j - \mathrm{H.c.} \right)
\end{aligned}
\end{equation}
where $n = 1$ or $2$ and $m \neq 1$ or $2$.
The QFI with respect to these states is straightforward to calculate, using the operator $\hat{A}_x^{(1,m)} = \left( \hat{a} \hat{c}^\dagger + \hat{c} \hat{a}^\dagger \right)/2$ where $\hat{a}$ is the annihilation operator for the mode $\ket{1}$, $\hat{b}$ is the annihilation operator for the mode $\ket{2}$, and $\hat{c}$ is the annihilation operator for mode $\ket{m}$, so that $\ket{\mathcal{N}} = (\ket{N,0,0} + \ket{0,N,0})/\sqrt{2}$.
This lets us calculate:
\begin{equation}
\begin{aligned}
\lambda =& 4 \left( \langle ( \hat{A}^{(1,m)}_x )^2 \rangle_{\mathcal{N}} - \langle \hat{A}^{(1,m)}_x \rangle_{\mathcal{N}}^2 \right) \\
=& 4 \langle (\hat{A}_x^{(1,m)})^2 \rangle_{\mathcal{N}} \\
=& \bra{\mathcal{N}}  ( \hat{a} \hat{c}^\dagger + \hat{c} \hat{a}^\dagger )^2 \ket{\mathcal{N}} \\
=& \frac{N}{2},
\end{aligned}
\end{equation}
where the $\langle \hat{A}_x \rangle_{\mathcal{N}} = 0$ and assuming that $N\geq2$. 
For $N=1$, we have $\ket{\mathcal{N}} = \ket{\mathcal{C}}$ are the same.
The same argument as in~\cref{eq:NoOffDiagonals} holds, whereupon we conclude that a Noon state has a QFIM with:
\begin{equation}
\begin{aligned}
& \to 1 \ \mathrm{eigenvalue\ of} \ N^2 \\
& \to 2 \ \mathrm{eigenvalues\ of} \ N \\
& \to 4(d-2) \ \mathrm{eigenvalues\ of} \ N/2 \\
\end{aligned}
\end{equation}
This yields
\begin{equation}
\begin{aligned}
\Tr_{\su{d}}(\mathbf{F}^{\su{d}}[\mathcal{N}]) &= N^2 + 2 N (d-2),\\
\mathrm{pdet}(\mathbf{F}^{\su{d}}[\mathcal{N}]) &= N^4\left(\frac{N}{2}\right)^{4(d-2)}.
\end{aligned}
\end{equation} 
We note something interesting, in the case that $N=2$, i.e. a Bell pair with $d$ possible levels, $\mathrm{pdet}(\mathbf{F}^{\g}[\mathcal{N}]) = 2^4$ is fixed while $\mathrm{pdet}(\mathbf{F}^{\g}[\mathcal{C}]) = 2^{2(d-1)}$ grows with respect to $d$.
This can be combined with the uncertainty principles given at the end of the main text to yield interesting implications about the usefulness of coherent states vs Noon states for multiparameter sensing.
This also motivates the fact that only the trace of the QFIM can be used as an inter-particle entanglement witness.
In fact, $ \mathrm{pdet}(\mathbf{F}^{\g}[\mathcal{N}]) = \mathrm{pdet}(\mathbf{F}^{\g}[\mathcal{C}])$ when $N = 2^{2 - \frac{2}{d-1}}$, which only has a valid integer solution at $N=2$ and $d=3$.

\subsection{Trace as an Entanglement Witness}
We have that $\Tr_{\g}(\mathbf{F}^{\g}[\mathcal{C}]) = 2 N (d-1)$ for a CSS.
First, we can verify that no separable state can surpass this through the following, where $\rho$ is pure:
\begin{equation}
\begin{aligned}
\Tr_{\g}( \mathbf{F}^\g[\rho] ) &= 4 \sum_\mu \langle \hat{G}_\mu^2 \rangle_\rho - \langle \hat{G}_\mu \rangle_\rho^2 \\
&= 4 \langle \hat{\mathcal{G}}^2 \rangle_\rho - 4 \sum_\mu \langle \hat{G}_\mu \rangle_\rho^2 \\
\end{aligned}
\end{equation}
where $\hat{\mathcal{G}^2}$ is the quadratic Casimir operator of $\su{d}$.
The maximal eigenvalue of $\hat{\mathcal{G}^2}$ is on the permutationally symmetric irreducible representation--i.e. when all the qudits are permutationally symmetric.
Since $ 4 \sum_\mu \langle \hat{G}_\mu \rangle_\rho^2$ is strictly positive, we know that a coherent spin state of the form in the main text has the largest trace QFIM for pures states.

Furthermore, no mixed state can surpass this:
Let
\begin{equation}
\rho_q = \frac{1}{q} \sum_{n=1}^q \left(\op{n}{n}\right)^{\otimes N}, \quad \mathrm{for} \ 1\leq q \leq d
\end{equation}
where $\ket{n}$ is a single particle state labeled in an arbitrary basis.
Then
\begin{equation}
\begin{aligned}
\Tr_{\g}(\mathbf{F}^{\su{d}}[\rho_q]) =& \sum_j \lambda_j[\rho_q] \\
\leq& \sum_j \left( \frac{1}{q} \sum_{n=1}^q \lambda_j\left[\left(\op{n}{n}\right)^{\otimes N}\right] \right)\\
=&  \left( \frac{1}{q} \sum_{n=1}^q \Tr_{\g}(\mathbf{F}^{\g}[\left(\op{n}{n}\right)^{\otimes N}]) \right)\\
=&  2 N(d-1).
\end{aligned}
\end{equation}
In order for the trace of the QFIM to surpass this scaling, one of the eigenvalues must scale faster than $N$.
This is an indication of entanglement~\cite{Toth}.
Therefore,
\begin{equation} \label{eq:TrCondition}
\Tr_{\g}(\mathbf{F}^{\g}[\rho]) \geq 2 N(d-1),
\end{equation} 
is an entanglement witness.

\section{The Average Response as the Trace QFIM}\label{sec:AvgRes}
In the main text we consider a small unitary applied to the probe state; $\hat{U}(\epsilon,\vec{n}) = \mathrm{e}^{-i \epsilon \hat{G}_{\vec{n}} }$, where $\epsilon\ll1$, $\vec{n}$ is a $\dim(\g)$ unit vector and $\hat{G}_{\vec{n}} = n^\mu \hat{G}_\mu \in \g$.
We state that the set of all these actions that form the $\epsilon$-ball of actions near the identity in the Lie group $\G$.
Formally speaking, we mean that the Lie group, $\G$, is a manifold and we are considering the set 
\begin{equation}
\mathcal{B}(\epsilon) = \{ \hat{U}(\epsilon,\vec{n}) \in \G | \epsilon \ll 1, |\vec{n}|^2 = 1 \}.
\end{equation}
The distance on $\G$ is given by the geodesic distance, but for $\epsilon \ll 1$ we can us the metric on the group--which is the Hilbert-Schmidt inner product in this case:
\begin{equation}
\begin{aligned}
\Delta_\G (\hat{U}(\epsilon,\vec{n}) , \hat{1} )^2 \approx& n^\mu \Tr_{\g}( \hat{G}_\mu, \hat{G}_\nu ) n^\nu \epsilon^2 \\
=& C n^\mu \delta_{\mu\nu} n^\nu \epsilon^2 = C \epsilon^2 ,
\end{aligned}
\end{equation}
where we used that $n^\mu \delta_{\mu\nu} n^\nu =  |\vec{n}|^2 = 1 $.
So, on $\G$ the $\epsilon$-ball is isotropic and has radius $r_\G = \epsilon \sqrt{C}$.

Now, in the main text we want to see how this ball deforms when it is included into $M_\rho$.
We can include this ball into the state space by the adjoint action:
\begin{equation}
\tilde{\rho}(\vec{n}) = \hat{U}(\epsilon,\vec{n}) \rho \hat{U}(-\epsilon,\vec{n}),
\end{equation}
where $\tilde{\rho}(\vec{n}) \in M_\rho$.
To see how it deforms, we derive the mean square distance from $\rho$ to $\tilde{\rho}(\vec{n})$, which find to be
\begin{equation}
\overline{r_\g^2} \equiv \epsilon^2 \ \frac{\Tr_{\g}(\mathbf{F}^{\g}[\rho])}{\dim(\g)} \frac{1}{8}
\end{equation}
which was found from the integral:
\begin{equation}
\begin{aligned}
\overline{r_\g^2} \equiv \frac{1}{A} \int_{S^{\dim(\g)}} \Delta^2 dA,
\end{aligned}
\end{equation}
where $A = ( 2 \pi^{\dim(\g)/2} ) / \Gamma( \frac{\dim(\g)}{2} ) $ is the surface area of the $\dim(\g)$-sphere, $\Gamma$ is the gamma function, and
\begin{equation}
\begin{aligned}
\Delta^2 =& 1 - \Tr_{\mh}\left(\sqrt{\sqrt{\rho} \tilde{\rho}(\epsilon,\vec{n}) \sqrt{\rho} }\right) \\
\approx& \ \vec{n}^T \mathbf{F}^\g[\rho] \vec{n} \ \frac{\epsilon^2}{8}.
\end{aligned}
\end{equation}
Here, we'll solve this integral explicitly. 
It reduces to a well known integral quickly.
Let $D = \dim(\g)$ for conciseness.
First, we use expansion of the Uhlmann distance according to small $\epsilon$:
\begin{equation}
\begin{aligned}
\frac{1}{A} \int_{S^D} \Delta^2 dA &= \frac{\epsilon^2}{8 A} \int_{S^{D}} \vec{n}^T \mathbf{F}^{\g}[\rho] \vec{n} dA \\
&= \frac{\epsilon^2}{8 A} \sum_{\mu,\nu} \mathcal{F}^{\g}[\rho]_{\mu\nu} \int_{S^{D}} n^\mu n^\nu dA.
\end{aligned}
\end{equation}
So, we have to evaluate the integral
\begin{equation}
\frac{1}{A} \int_{S^{D}} n^\mu n^\nu dA,
\end{equation}
which is a well known integral in probability and statistics.
To integrate it, we'll use hyper spherical coordinates.
This is $D-1$ many angles, $\theta^j$ such that
\begin{equation}
\begin{aligned}
n^1 =& \cos(\theta^1), \\
n^2 =& \sin(\theta^1)\cos(\theta^2), \\
\vdots \\
n^{D-1} =& \sin(\theta^1) \sin(\theta^2) \cdots \sin(\theta^{D-2}) \cos(\theta^{D-1}),\\
n^{D} =& \sin(\theta^1) \sin(\theta^2) \cdots \sin(\theta^{D-2}) \sin(\theta^{D-1})
\end{aligned}
\end{equation}
where $\theta^{D-1} \in [0, 2 \pi]$ and $\theta^j \in [0,\pi]$ for $j = 1,\dots,D-2$.
The hyper spherical area element is 
\begin{equation}
dA =  \prod_{j=1}^{D-1} \sin(\theta^j)^{D-1-j} d\theta^{j}.
\end{equation}
We can evaluate this by symmetry arguments.
On the $D$-sphere, $n^\mu$ has a minimum value of $-1$ and a maximum of $+1$, and is odd.
Therefore, differing coordinates $\mu\neq\nu$ will go to zero.
Now, the integral is spherically symmetric, so we know that all coordinates must contribute the same real number, call it $B$.
Therefore,
\begin{equation}
\frac{1}{A} \int_{S^{D}} n^\mu n^\nu dA = B \delta^{\mu\nu},
\end{equation}
where we have to determine $B$. We can do this by choosing the simplest coordinate:
\begin{equation}
\begin{aligned}
\int_0^{\pi} \cos(\theta^1)^2 \sin(\theta^1)^{D-2} d\theta^1 = \frac{\sqrt{\pi} \Gamma\left(\frac{D-1}{2}\right)}{2 \Gamma\left(\frac{D+2}{2}\right)}.
\end{aligned}
\end{equation}
Unless $D=2$, in which case this is integration around the unit circle and left as an exercise for the reader.
This leaves the $D-1$ angle which integrates to $2 \pi$, and $D-3$ other angles of the form
\begin{equation}
\int_0^{\pi} \sin(\theta^j)^{D-1-j} d\theta^j = \frac{\sqrt{\pi} \Gamma\left( \frac{D-j}{2} \right)}{\Gamma\left(\frac{D - j+1}{2} \right) }
\end{equation}
We can then directly calculate $B$ where, luckily, many terms cancel
\begin{equation}
\begin{aligned}
B =& \frac{2 \pi }{A} \frac{\sqrt{\pi} \Gamma\left(\frac{D-1}{2}\right)}{2 \Gamma\left(\frac{D+2}{2}\right)} \prod_{j=2}^{D-2} \left(  \frac{\sqrt{\pi} \Gamma\left( \frac{D-j}{2} \right) }{ \Gamma\left(\frac{D - j+1}{2} \right) }\right) \\
=& \frac{1}{2} \frac{\Gamma\left(\frac{D}{2}\right) }{\Gamma\left(\frac{D+2}{2}\right)} = \frac{1}{D},
\end{aligned}
\end{equation}
as stated in the main text.
Thus, all the steps summarized are
\begin{equation}
\begin{aligned}
\frac{1}{A} \int_{S^D} \Delta^2 dA &= \frac{\epsilon^2}{8 A} \int_{S^{D}} \vec{n}^T \mathbf{F}^{\g}[\rho] \vec{n} dA \\
&= \frac{\epsilon^2}{8 A} \sum_{\mu,\nu} \mathcal{F}^{\g}[\rho]_{\mu\nu} \delta^{\mu\nu} \frac{1}{\dim(\g)} \\
&= \frac{\epsilon^2}{8} \frac{\Tr_{\g}(\mathbf{F}^{\g}[\rho])}{\dim(\g)} .
\end{aligned}
\end{equation}

\section{The Model for the Illustrative Example}\label{sec:Model}

\begin{figure}[h]
    \centering
    \includegraphics[width=1 \columnwidth]{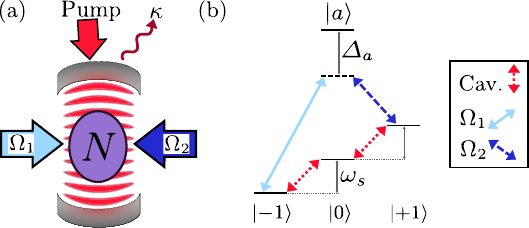}
    \caption{
    (a) A cartoon of the illustrative example.
    (b) The expanded level structure of the atoms, including the far detuned auxiliary state which provides the virtual level (dotted lines) for a two-photon interactions.}
    \label{fig:Fig_S1}
\end{figure}

Here, we provide a brief derivation of the physical model used in the example in the main text.
We consider $N$ atoms in a cavity, and three lasers.
The first laser shines on the cavity mode, and the other two directly on the atoms.
The cavity and pump scheme is shown in~\cref{fig:Fig_S1} (a).

The atoms have three relevant states labeled $\ket{\pm1}$ and $\ket{0}$ that are split by $\omega_s$.
To be clear, the purpose of this model is to recover the two simple unitary interactions.
The simplest way to find this level structure is the hyperfine levels of a ground state manifold of an atom, with energy levels that can be tuned by a magnetic field--however polar molecules or atoms with an excited and ground state manifold and multi-photon processes for spin squeezing would work with optical wavelength cavities.
These atoms have one far detuned axuiliary state, $\ket{a}$, split by a large frequency $\omega_a$ to provide virtual state.
The cavity has one relevant mode, with frequency $\omega_c$ and decay rate $\kappa$, and is directly driven by a pump laser, at frequency $\omega_p$ and amplitude $\eta$.
Two lasers that are in phase with each other and the cavity are directly shined on the atoms, with amplitudes $\Omega_1$ and $\Omega_2$ and frequencies $\omega_1$ and $\omega_2$.
The cavity light drives transitions between adjacent states $\ket{-1} \leftrightarrow \ket{0} \leftrightarrow \ket{+1}$, while the two lasers directly illuminate the atoms to drive from $\ket{-1} \leftrightarrow \ket{a} \leftrightarrow \ket{+1}$.
The direct-drive lasers are never on at the same time the cavity is being pumped, and are far detuned from driving $\ket{\pm1} \leftrightarrow \ket{0}$.
This gives the energy structure and couplings shown in~\cref{fig:Fig_S1} (b).

In the main text, we set $\kappa = 0$ to explore just the unitary dynamics. 
For sake of completeness, we will leave it finite but set it to $0$ in the last step.
This means the density matrix evolves according to the Lindblad master equation
\begin{equation}
\frac{\partial \rho}{\partial t} = \mathcal{L} \rho = -i [ \hat{H}, \rho ] + \mathcal{D}[\sqrt{\kappa} \hat{a} ] \rho
\end{equation}
where $\mathcal{D}[\hat{O}] \rho = \hat{O} \rho \hat{O}^\dagger - \{\hat{O}^\dagger \hat{O}, \rho \}/2$ accounts for decoherence.
First, just the cavity pump is turned on to seed One-Axis Twisting (OAT).
The Hamiltonian is, in the interaction picture of the pump frequency, the spin-1 Jaynes-Cummings Hamiltonian:
\begin{equation}
    \begin{aligned}
        \hat{H}_{a,c}=& \Delta_c \hat{a}^\dag \hat{a}+\eta \left( \hat{a}^\dag + \hat{a} \right)
        \\& + \sum_{j=1}^N \Delta_s \left(\ket{+1}_j\bra{+1}_j-\ket{-1}_j\bra{\-1}_j\right) 
        \\& g\hat{a}^\dag \left( \ket{-1}_j\bra{0}_j + \ket{0}_j \bra{+1}_j \right) + \mathrm{H.c}.
    \end{aligned}
\end{equation}
where we may ignored $\ket{a}$ for now as it is decoupled. 
The detunings are $\Delta_c=\omega_c-\omega_p$ and $\Delta_s=\omega_s - \omega_p$.
We can rewrite this in terms of the operators in $\g_1$, where
\begin{equation}
    \begin{aligned}
        \hat{H}_{a,c}=& \hat{H}_\mathrm{cav} + \Delta_s \hat{S}_z + \frac{g}{\sqrt{2}} \left( \hat{a}^\dag \hat{S}^+ + \hat{S}^- \hat{a} \right) ,
    \end{aligned}
\end{equation}
with $\hat{C}_\mathrm{cav} = \Delta_c \hat{a}^\dag \hat{a}+\eta \left( \hat{a}^\dag + \hat{a} \right)$
We now eliminate the cavity based on large detuning, $|\Delta_c| \gg \eta, |\Delta_s|, g \langle \hat{a}^\dagger \hat{a} \rangle, \kappa \langle \hat{a}^\dagger \hat{a} \rangle.$

We eliminate the cavity following Ref.~\cite{Jager2022}.
Here, $\hat{H}_\mathrm{sys} = \Delta_s \hat{S}_z $, while $\eta + (g/\sqrt{2}) \hat{S}^+$ acts as the effective pump terms to the cavity elimination.
We solve
\begin{equation}~\label{eq:alpha}
\begin{aligned}
\frac{\partial \hat{\alpha}}{\partial t} = -i \Delta_s [ \hat{S}_z,\hat{\alpha}] &- i \Delta_c \hat{\alpha} - \kappa \hat{\alpha} - i \eta - i \frac{g}{\sqrt{2}} \hat{S}^+
\end{aligned}
\end{equation}
whereupon the effective evolution is given by
\begin{equation}
\frac{\partial \rho}{\partial t} = \mathcal{L}_\mathrm{eff} \rho = -i [ \hat{H}_a, \rho ] + \mathcal{D}[\sqrt{\kappa} \hat{\alpha} ] \rho,
\end{equation}
with the effective Hamiltonian being
\begin{equation}
\hat{H}_S = \hat{H}_\mathrm{sys} + \frac{1}{2} \left( \left( \eta + \frac{g}{\sqrt{2}} \hat{S}^+ \right) \hat{\alpha} + \mathrm{H.c.} \right).
\end{equation}
We can solve for $\hat{\alpha}$ from~\cref{eq:alpha} to find
\begin{equation}
\hat{\alpha} = \frac{\eta}{-\Delta_c + i \kappa } -\frac{g}{\sqrt{2} (\Delta - i \kappa)} \hat{S}^-
\end{equation}
where we have defined $\Delta = \omega_c - \omega_s $ as the cavity-hyperfine splitting.
Here, we will take $\kappa = 0$, which is not physical but we want to explore the conservation of a simple unitary model.
The effective Hamiltonian is
\begin{equation}
\begin{aligned}
\hat{H}_S =& \Delta_s \hat{S}_z + \frac{g \eta}{\sqrt{2}} \left( \frac{1}{ \Delta} -\frac{1}{\Delta_c} \right) \hat{S}_x + \frac{g^2}{2 \Delta} \hat{S}^+\hat{S}^- .
\end{aligned}
\end{equation}
Here, we can consider a small drive to shut off the $\hat{S}_x$ rotation, and expand these terms to find
\begin{equation}
\begin{aligned}
\hat{H}_S =& \Delta_s \hat{S}_z + \chi \hat{S}^2 - \chi \hat{S}_z^2 + \chi \hat{S}_z ,
\end{aligned}
\end{equation}
where $\chi = g^2/2\Delta$.
For the purpose of our mdel, we will shift the hyperfine spitting such that $\Delta_s = \chi$. 
However, one could optimize this.
This yields the final Hamiltonian
\begin{equation}
\begin{aligned}
\hat{H}_S =& \chi \hat{S}^2 - \chi \hat{S}_z^2.
\end{aligned}
\end{equation}
Here, we make an important point.
Because $\g_1$ is a reducible representation of $\su{2}$, $\hat{S}^2$ \textit{is not constant}.
However, we consider a starting state which is an eigenstate of $\hat{S}^2$, so we ignore it in the main text.

This Hamiltonian is applied for a time $t_s$, before the two pumps are turned on at a strength $\Omega_1, \Omega_2 \gg g \langle \hat{a}^\dagger \hat{a} \rangle$, so we ignore the cavity in this second step, and $\omega_s$ need not be tuned to cancel the extra $\hat{S}_z$ rotation here.
The Hamiltonian in the interaction frame of the two laser frequencies is,
\begin{equation}
    \begin{aligned}
        \hat{H}_{a,L}= \Delta_s \hat{S}_z + \Delta_a \hat{N}_a  &+\frac{\hbar\Omega_1}{2}\left( |a\rangle_j\langle +1|_j +|+1\rangle_j\langle a|_j \right)
        \\&+\frac{\hbar\Omega_2}{2} \left( |a\rangle_j\langle -1|_j +|-1\rangle_j\langle a|_j \right) 
    \end{aligned}
\end{equation}
where $\hat{N}_a = \sum_{j=1}^N \ket{a}_j \bra{a}_j$ and $\Delta_s=\frac{-\omega_1+\omega_2}{2}-\omega_s$ and $ \Delta_a=\frac{\omega_1+\omega_2}{2}-\omega_a$.
We tune $\omega_s$ such that $\Delta_s = 0$ and subsequently set $\Omega_1 = \Omega_2$, $\omega_1 = \omega_2$.
We eliminate the excited state based on $|\Delta_a| \gg \Omega_1$, and find
\begin{equation}
    \begin{aligned}
    \hat{H}_{R} =& 
        \frac{\Omega_1^2}{4 \Delta_a} \hat{N}_{0} + \Omega \hat{Q}_3
    \end{aligned}
\end{equation}
where $\Omega = \Omega_1 \Omega_2/ 2 \Delta_a$ and $\hat{N}_m = \sum_{j=1}^N \ket{m}_j \bra{m}_j$.
We used the fact that $\hat{N}_{+1} + \hat{N}_{-1} = N - \hat{N}_0$ and dropped the constant terms.
In the main text we ignored the term $\Omega_1^2 \hat{N}/\Delta_a$, which does not change the application of the theorem since $\hat{N}_0$ can be expressed in $\g_2$.

\end{document}

%% file: main.bbl
%

%% file: Geometric_Invariants_of_Quantum_Metrology.bbl
\begin{thebibliography}{95}
\bibitem [{Note7({\natexlab{a}})}]{footnoteMetric}%
  \BibitemOpen
  \bibinfo {note} {In index notation, we'd say that a matrix must have one upper and one lower index. This is because a matrix, $\mathbf{M}$, maps a vector to a vector; $M^{\alpha}_\beta V^\beta = W^\alpha$.}\BibitemShut {Stop}%
\bibitem [{Note7({\natexlab{b}})}]{footnoteHaar}%
  \BibitemOpen
  \bibinfo {note} {We note that this is equivalent to integrating against the Haar measure on $\G$ to pick a random direction for $\hat{G}_{\vec{n}} = \hat{V}(\vec{n}) \hat{G}_1 \hat{V}(\vec{n})^\dagger$ where $\hat{V}(\vec{n}) \in \G$ and $\hat{G}_1$ is the (arbtirarily chosen) first element of our basis for $\g$.}\BibitemShut {Stop}%
\bibitem [{Note6()}]{footnoteSubalgebras}%
  \BibitemOpen
  \bibinfo {note} {This could equivalently be found by the observation that $\su{d}$ has a maximal Cartan sub-algebra of dimension $d-1$, which yields a counting argument that there are $2(d-1)$ unique directions of rotation.}\BibitemShut {Stop}%
\end{thebibliography}
